\documentclass[acmsmall]{acmart}

\AtBeginDocument{%
  }

\usepackage{tikz}
\usetikzlibrary{mindmap}
\usepackage{metalogo}
\usepackage{smartdiagram}
\usesmartdiagramlibrary{additions}
\usepackage{forest}
\usetikzlibrary{shadows}
\usepackage{cleveref}
\usepackage{booktabs}
\usepackage{multirow}
\usepackage{makecell}
\usepackage{subfigure}
\usepackage[misc]{ifsym} 
\usepackage{balance}
\usepackage{listings}
\usepackage{color, xcolor}
\usepackage[linesnumbered,ruled,vlined]{algorithm2e}
\usepackage{tcolorbox}
\usepackage{amsmath}

\definecolor{lightcoral}{rgb}{0.94, 0.5, 0.5}
\definecolor{lightgreen}{rgb}{0.56, 0.93, 0.56}
\definecolor{harvestgold}{rgb}{0.85, 0.57, 0.0}
\definecolor{brightlavender}{rgb}{0.75, 0.58, 0.89}
\definecolor{capri}{rgb}{0.0, 0.75, 1.0}
\definecolor{carminepink}{rgb}{0.92, 0.3, 0.26}
\definecolor{celadon}{rgb}{0.67, 0.88, 0.69}
\definecolor{darkpastelgreen}{rgb}{0.01, 0.75, 0.24}
\definecolor{DeepSkyBlue4}{RGB}{0,104,139}

\newcommand{\blueref}[1]{\textcolor{blue}{\ref{#1}}}

\setcopyright{acmlicensed}
\copyrightyear{2024}
\acmYear{2024}
\acmDOI{XXXXXXX.XXXXXXX}

\acmJournal{JACM}
\acmVolume{37}
\acmNumber{4}
\acmArticle{111}
\acmMonth{8}

\begin{document}

\title{A Survey of AIOps for Failure Management in the Era of Large Language Models}

\author{Lingzhe Zhang}
\affiliation{%
	\institution{Peking University}
	\city{Beijing}
	\country{China}}
\email{zhang.lingzhe@stu.pku.edu.cn}

\author{Tong Jia$^{\ast}$}
\thanks{*Corresponding author}
\affiliation{%
	\institution{Peking University}
	\city{Beijing}
	\country{China}}
\email{jia.tong@pku.edu.cn}

\author{Mengxi Jia}
\affiliation{%
	\institution{Peking University}
	\city{Beijing}
	\country{China}}
\email{mxjia@pku.edu.cn}

\author{Yifan Wu}
\affiliation{%
	\institution{Peking University}
	\city{Beijing}
	\country{China}}
\email{yifanwu@pku.edu.cn}

\author{Aiwei Liu}
\affiliation{%
	\institution{Tsinghua University}
	\city{Beijing}
	\country{China}}
\email{liuaw20@mails.tsinghua.edu.cn}

\author{Yong Yang}
\affiliation{%
	\institution{Peking University}
	\city{Beijing}
	\country{China}}
\email{yang.yong@pku.edu.cn}

\author{Zhonghai Wu}
\affiliation{%
	\institution{Peking University}
	\city{Beijing}
	\country{China}}
\email{wuzh@pku.edu.cn}

\author{Xuming Hu}
\affiliation{
	\institution{The Hong Kong University of Science and Technology (Guangzhou)}
	\city{Guangzhou}
	\country{China}}
\email{xuminghu@hkust-gz.edu.cn}

\author{Philip S. Yu}
\affiliation{
	\institution{University of Illinois Chicago}
	\city{Chicago}
	\country{United States}}
\email{psyu@cs.uic.edu}

\author{Ying Li$^{\ast}$}
\affiliation{%
	\institution{Peking University}
	\city{Beijing}
	\country{China}}
\email{li.ying@pku.edu.cn}

\renewcommand{\shortauthors}{Lingzhe Zhang et al.}

\begin{abstract}
  As software systems grow increasingly intricate, Artificial Intelligence for IT Operations (AIOps) methods have been widely used in software system failure management to ensure the high availability and reliability of large-scale distributed software systems. However, these methods still face several challenges, such as lack of cross-platform generality and cross-task flexibility. Fortunately, recent advancements in large language models (LLMs) can significantly address these challenges, and many approaches have already been proposed to explore this field. However, there is currently no comprehensive survey that discusses the differences between LLM-based AIOps and traditional AIOps methods. Therefore, this paper presents a comprehensive survey of AIOps technology for failure management in the LLM era. It includes a detailed definition of AIOps tasks for failure management, the data sources for AIOps, and the LLM-based approaches adopted for AIOps. Additionally, this survey explores the AIOps subtasks, the specific LLM-based approaches suitable for different AIOps subtasks, and the challenges and future directions of the domain, aiming to further its development and application.
\end{abstract}

\ccsdesc[500]{Software and its engineering~Maintaining software}
\ccsdesc[500]{Computing methodologies~Artificial intelligence}
\ccsdesc[500]{General and reference~Surveys and overviews}

\keywords{Large Language Model, AIOps, Failure Management, Metrics, Logs, Time Series, Failure Perception, Anomaly Detection, Root Cause Analysis, Auto Remediation}


\maketitle

\section{Introduction}

Nowadays, software systems are becoming increasingly complex. These systems typically serve massive user bases numbering in the billions, where even minor software glitches can result in significant losses due to service interruptions or degraded service quality~\cite{elliot2014devops, ma2020diagnosing}. Therefore, large-scale distributed software systems need to ensure uninterrupted 24/7 service, with high availability and reliability requirements. However, due to their vast scale and intricate logic of operation, these software systems frequently experience failures that are challenging to detect, pinpoint, and diagnose. Analyzing and debugging system errors become even more difficult once faults occur. Thus, enhancing fault diagnosis efficiency, swiftly identifying system failures, pinpointing root causes, and promptly remedying them have become critical for ensuring the high availability and reliability of large-scale distributed software systems.

With the advancement of Artificial Intelligence (AI), Artificial Intelligence for IT Operations (AIOps) was first introduced by Gartner in 2016~\cite{prasad2018market}. AIOps leverages machine learning (ML) or deep learning (DL) algorithms to analyze vast amounts of data from various operational tools and devices, automatically detecting and responding to system issues in real-time. This enhances the capabilities and automation levels of Information Technology (IT) operations. Consequently, AIOps for failure management (FM) has become a mainstream approach to ensuring high availability and reliability of software systems.

\subsection{Why are LLMs Beneficial for AIOps on Failure Management?}
\label{sec: llm4aiops}

While aformethioned ML-based or DL-based AIOps methods for failure management have significantly assisted in software system operations, they still face several challenges as follows:

\begin{itemize}
	\item \textbf{Need for complex feature extraction engineering.} These AIOps methods typically require extensive preprocessing and feature engineering to extract useful information from raw data. They have limited capabilities in understanding and processing data, especially in handling unstructured data such as logs and traces, which appearing relatively weak.
	\item \textbf{Lack of cross-platform generality.} Traditional AIOps models are often tuned and trained specifically for a particular software system. Once a different software system is adopted or even minor changes are made to the original system, the performance of the model significantly deteriorates, even when performing the same task.
	\item \textbf{Lack of cross-task flexibility.} Due to the singularity of model knowledge and outputs, AIOps models can only perform one task at a time. For example, in Root Cause Analysis (RCA) tasks, some work is aimed at identifying the cause of the problem~\cite{sui2023logkg, yuan2019approach, lin2016log}, while others are focused on identifying the software components involved~\cite{misiakos2024learning, ikram2022root, wang2023root}. In real-world scenarios, multiple models must run simultaneously to complete the entire RCA task.
	\item \textbf{Limited model adaptability.} With changes in the software system, deep learning-based AIOps methods typically require frequent model training and updates to adapt to new data and environments. While there are many online learning methods~\cite{ahmed2021anomaly, lyu2023assessing, li2023few, han2021log} available to address this issue, this process not only consumes time and effort but also can result in delayed responses from the model when handling sudden events.
	\item \textbf{Restricted levels of automation.} Current deep learning-based AIOps methods exhibit relatively limited capabilities in terms of automated operations and intelligent decision-making. While some degree of automation is achievable, significant manual intervention and configuration are still required. Particularly in the case of Auto Remediation, current efforts are mainly stopped at Incident Triage~\cite{shao2008efficient, zeng2017knowledge} or Solution Recommendation~\cite{lin2018hardware, wang2017constructing, zhou2016resolution}.
\end{itemize}

Large Language Models (LLMs), pre-trained on natural language understanding tasks, offer a promising avenue for addressing these limitations. (1) Due to their robust natural language processing capabilities, LLMs can efficiently handle and comprehend unstructured data, often eliminating the need for prior feature extraction. (2) Trained on vast amounts of cross-platform data, LLMs possess a strong degree of generality and logical reasoning abilities. (3) Outputting natural language, LLMs offer great flexibility, enabling them to simultaneously perform multiple AIOps tasks, such as identifying the cause of a problem and the involved software components. (4) Leveraging their pre-training, LLMs exhibit powerful adaptive capabilities and can incorporate continuously updated external knowledge using methods like Retrieval-Augmented Generation (RAG), often without requiring retraining. (5) With robust script generation capabilities and the ability to automatically invoke external tools, LLMs can achieve higher levels of automation.

\begin{table}[htbp]
	\centering
	\caption{State-of-art survey related to AIOps for failure management (FM)}
	\label{tab: survey-comparison}
	\begin{tabular}{ccccc}
		\toprule
		Reference & Year & Scope of AIOps tasks & Scope of FM tasks & LLM-based \\
		\midrule
		Paolop et al.~\cite{notaro2021survey} & 2021 & Failure Management & \makecell[c]{Failure Perception\\Root Cause Analysis\\Remediation} &  \\
		\midrule
		Josu et al.~\cite{diaz2023joint} & 2023 & \makecell[c]{Challenges, Architectures,\\ Future Fields} & \makecell[c]{Anomaly Detection\\Root Cause Analysis\\Auto Remediation} &  \\
		\midrule
		Angela et al.~\cite{fan2023large} & 2023 & \makecell[c]{Requirement Engineering\\Design \& Planning\\Code \& Testing\\Maintainance \& Deployment} & - & $\checkmark$ \\
		\midrule
		Qian et al.~\cite{cheng2023ai} & 2023 & Failure Management & \makecell[c]{Incident Detection\\Failure Prediction\\Root Cause Analysis\\Automated Actions} &  \\
		\midrule
		Zhang et al.~\cite{zhang2023survey} & 2023 & \makecell[c]{Code-related tasks\\Requirements \& Design\\Software Development\\Software Testing\\Software Maintenance} & - & $\checkmark$ \\
		\midrule
		Youcef et al.~\cite{remil2024aiops} & 2024 & Incident Management & \makecell[c]{Incident Reporting\\Incident Triage\\Incident Diagnosis\\Incident Mitigation} &  \\
		\midrule
		Wei et al.~\cite{wei2024log} & 2024 & Failure Management & Anomaly Detection &  \\
		\midrule
		Jing et al.~\cite{su2024large} & 2024 & Failure Management & \makecell[c]{Time-series Forecasting\\Anomaly Detection} & $\checkmark$ \\
		\midrule
		\textbf{Our work} & - & Failure Management & \makecell[c]{Data Preprocessing\\Failure Perception\\Root Cause Analysis\\Auto Remediation} & $\checkmark$ \\
		\bottomrule
	\end{tabular}
\end{table}

\subsection{Why a Survey of AIOps for Failure Management in the Era of LLMs? }

Numerous literature reviews have summarized research on AIOps. As shown in Table~\ref{tab: survey-comparison}, these works are either based on traditional machine learning or deep learning algorithms and do not use LLM-based approaches~\cite{notaro2021survey, cheng2023ai, remil2024aiops, wei2024log}, or they do not provide a systematic summary of all tasks involved in the full process of AIOps for failure management~\cite{su2024large}. Some reviews may not even specifically focus on failure management within the realm of AIOps~\cite{diaz2023joint, fan2023large, zhang2023survey}.

In summary, comprehensive studies exploring AIOps for failure management in the era of LLMs are lacking. However, as illustrated in Section~\ref{sec: llm4aiops}, LLM-based approaches offer significant benefits for AIOps tasks. In this study, we present the first comprehensive survey that covers the entire process of AIOps for failure management in the context of large language models. This survey encompasses a detailed definition of AIOps tasks for failure management, the data sources for AIOps, and the LLM-based approaches adopted for AIOps. Moreover, we delve into the AIOps subtasks and the specific LLM-based approaches suitable for different AIOps subtasks. This survey aims to provide researchers with an in-depth understanding of LLM-based approaches for AIOps, facilitating comparisons and contrasts among different methods. It also guides users interested in applying LLM-based AIOps methods by helping them choose suitable algorithms for different application scenarios.

\textbf{Organization of this survey. } This survey is structured as follows: Section 2 introduces the necessary preliminaries, including data sources for AIOps tasks, LLM-based approaches for AIOps, and AIOps tasks for failure management. Sections 3 through 6 detail the new characteristics and methods of four types of AIOps tasks in the era of LLMs according to a specific taxonomy in AIOps tasks: data preprocessing, failure perception, root cause analysis, and auto remediation. Section 7 examines ongoing challenges and potential future research avenues in LLM-based AIOps for failure management. The survey concludes in Section 8.

\section{Preliminary}

To facilitate the introduction of various AIOps tasks and their LLM-based methods in subsequent sections, this section presents the data sources adopted for AIOps, the LLM-based approaches widely used in AIOps, and the taxonomy of AIOps tasks for failure management.

\subsection{Data Source for AIOps}
\label{sec:data-source}

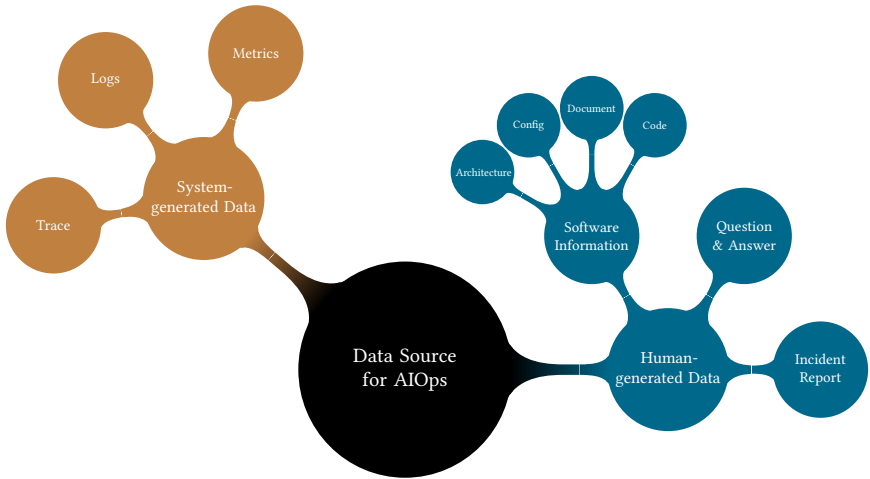
\begin{figure}[htbp]
	\centering
	\begin{tikzpicture}[scale=0.7, transform shape]
		\path[mindmap,concept color=black,text=white]
		node[concept] {Data Source for AIOps} [clockwise from=45]
		child[concept color=DeepSkyBlue4,grow=0]{
			node[concept] {Human-generated Data} [clockwise from=120]
			child { node[concept] {Software Information} [clockwise from=150]
				child{node[concept] {Architecture}}
				child{node[concept] {Config}}
				child{node[concept] {Document}}
				child{node[concept] {Code}}
			}
			child { node[concept] {Question \& Answer}} 
			child { node[concept] {Incident Report} }
		}
		child[concept color=brown,grow=140]{ 
			node[concept] {System-generated Data} [counterclockwise from=70]
			child {node[concept] {Metrics}}
			child {node[concept] {Logs}}
			child {node[concept] {Trace}}
		};
	\end{tikzpicture}
	\caption{Data Source of AIOps for Failure Management}
\label{fig: data-source}
\end{figure}

As shown in Table~\ref{fig: data-source}, the data sources adopted for AIOps can be divided into two categories based on their origin: system-generated data and human-generated data. System-generated data is automatically produced by the system, whereas human-generated data is created by people, including developers, operations personnel, and even users.

\textbf{System-generated Data.} This type of data is the most commonly used in traditional ML-based and DL-based AIOps for failure management~\cite{li2021opengauss, zhou2021dbmind, zhou2018fault, sui2023logkg, yuan2019approach, lin2016log, zhang2022deeptralog, zeng2023traceark, zhang2022putracead}. The data types include \textbf{metrics}, \textbf{logs}, and \textbf{traces}.

\begin{itemize}
	\item \textbf{Metrics.} Metrics are quantitative measurements collected from various components of the IT infrastructure, such as CPU usage, memory usage, disk I/O, network latency, and throughput. They provide real-time performance data and are crucial for monitoring the health and performance of the system.
	\item \textbf{Logs.} Logs are detailed records of events that occur within the system. They can include error messages, transaction records, user activities, and system operations. Logs are essential for diagnosing issues, understanding system behavior, and tracking changes over time.
	\item \textbf{Trace.} Traces are records of the sequence of operations or transactions that a request goes through in a distributed system. They provide a high-level view of how different services interact and help identify performance bottlenecks, dependencies, and the root cause of issues in microservices architectures.
\end{itemize}

\textbf{Human-generated Data.} With the emergence of LLMs, in addition to system-generated data, many approaches also utilize information created by humans to provide auxiliary knowledge for software system failure management~\cite{bhavya2023exploring, zhang2024automated, goel2024x, cao4741492managing, zhang2023pace, xue2023db, guo2023owl, roy2024exploring, shi2023shellgpt, ahmed2023recommending, jiang2024xpert, hamadanian2023holistic, jin2023assess}. The data types include \textbf{Software Information}, \textbf{Question \& Answer (QA)}, and \textbf{Incident Reports}.

\begin{itemize}
	\item \textbf{Software Information.} Software information is generated during the software development process and includes details such as \textit{software architecture}, \textit{configurations}, \textit{documentation}, and \textit{implementation code}. This type of information provides valuable knowledge about the development process, helping AIOps approaches to better perform various failure management tasks. By leveraging software information, AIOps can gain insights into the system's design and functionality, which can be crucial for diagnosing issues and implementing effective solutions.
	\item \textbf{Question \& Answer (QA).} QA data consists of question-and-answer pairs related to operational or development knowledge. In LLMs, utilizing a rich repository of QA data can serve as a knowledge base that functions like a search engine, providing valuable assistance to operations personnel. Additionally, similar to software information, QA data can supply auxiliary knowledge to enhance AIOps solutions.
	\item \textbf{Incident Reports.} Incident reports are often written by software users. When an incident is created, the author specifies a title for the incident and describes relevant details such as error messages, anomalous behavior, and other information that could help with resolution. Before the emergence of LLMs, these incident reports were submitted to on-call engineers (OCEs) for diagnosis. However, due to the powerful natural language processing capabilities of LLMs, many approaches now automatically analyze these incident reports, diagnose faults, and even suggest mitigation steps. This automation enhances the efficiency of failure management by quickly identifying and addressing issues based on the detailed information provided in the reports.
\end{itemize}

\subsection{LLM-based Approaches for AIOps}

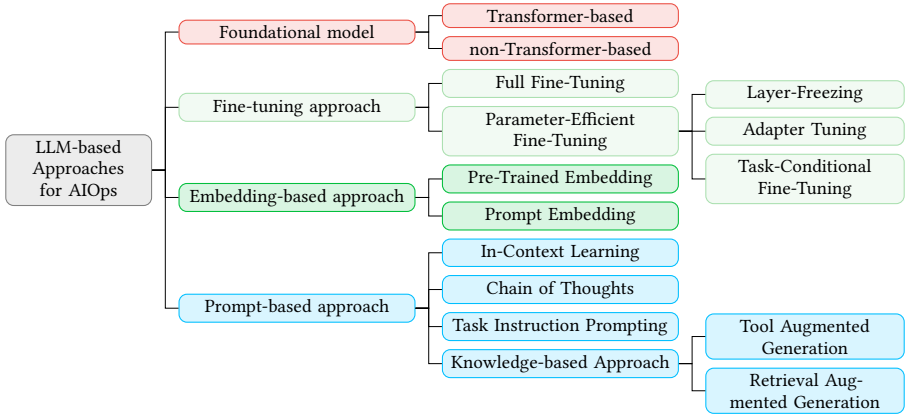
\begin{figure}[t]
	\centering
	\tikzset{
		my node/.style={
			draw,
			align=center,
			thin,
			text width=1.2cm, 
			rounded corners=3,
		},
		my leaf/.style={
			draw,
			align=left,
			thin,
			text width=8.5cm, 
			rounded corners=3,
		}
	}
	\forestset{
		every leaf node/.style={
			if n children=0{#1}{}
		},
		every tree node/.style={
			if n children=0{minimum width=1em}{#1}
		},
	}
	\begin{forest}
		nonleaf/.style={font=\scriptsize},
		for tree={%
			every leaf node={my node, font=\scriptsize, l sep-=4.5pt, l-=1.pt},
			every tree node={my node, font=\scriptsize, l sep-=4.5pt, l-=1.pt},
			anchor=west,
			inner sep=2pt,
			l sep=10pt, 
			s sep=3pt, 
			fit=tight,
			grow'=east,
			edge={ultra thin},
			parent anchor=east,
			child anchor=west,
			if n children=0{}{nonleaf}, 
			edge path={
				\noexpand\path [draw, \forestoption{edge}] (!u.parent anchor) -- +(5pt,0) |- (.child anchor)\forestoption{edge label};
			},
			if={isodd(n_children())}{
				for children={
					if={equal(n,(n_children("!u")+1)/2)}{calign with current}{}
				}
			}{}
		}
		[LLM-based \\Approaches for AIOps, draw=gray, fill=gray!15, text width=1.8cm, text=black
		[Foundational model, color=carminepink, fill=carminepink!15, text width=3cm, text=black
		[Transformer-based, color=carminepink, fill=carminepink!15, text width=3cm, text=black]
		[non-Transformer-based, color=carminepink, fill=carminepink!15, text width=3cm, text=black]
		]
		[Fine-tuning approach, color=celadon, fill=celadon!15, text width=3cm, text=black
		[Full Fine-Tuning, color=celadon, fill=celadon!15, text width=3cm, text=black]
		[Parameter-Efficient Fine-Tuning , color=celadon, fill=celadon!15, text width=3cm, text=black
		[Layer-Freezing, color=celadon, fill=celadon!15, text width=2.5cm, text=black]
		[Adapter Tuning, color=celadon, fill=celadon!15, text width=2.5cm, text=black]
		[Task-Conditional Fine-Tuning, color=celadon, fill=celadon!15, text width=2.5cm, text=black]
		],
		]
		[Embedding-based approach, color=darkpastelgreen, fill=darkpastelgreen!15, text width=3cm, text=black
		[Pre-Trained Embedding, color=darkpastelgreen, fill=darkpastelgreen!15, text width=3cm, text=black]
		[Prompt Embedding, color=darkpastelgreen, fill=darkpastelgreen!15, text width=3cm, text=black]
		]
		[Prompt-based approach, color=capri, fill=capri!15, text width=3cm, text=black
		[In-Context Learning, color=capri, fill=capri!15, text width=3cm, text=black]
		[Chain of Thoughts, color=capri, fill=capri!15, text width=3cm, text=black],
		[Task Instruction Prompting, color=capri, fill=capri!15, text width=3cm, text=black]
		[Knowledge-based Approach, color=capri, fill=capri!15, text width=3cm, text=black
		[Tool Augmented Generation, color=capri, fill=capri!15, text width=2.5cm, text=black]
		[Retrieval Augmented Generation, color=capri, fill=capri!15, text width=2.5cm, text=black]
		]
		]
		]
	\end{forest}
	\caption{LLM-based Approaches for AIOps}
	\label{fig: llm-approach}
\end{figure}

Many surveys have proposed various LLM-based approaches~\cite{li2024pre, chang2024survey, minaee2024large, gu2023systematic, sahoo2024systematic}, many of which have been applied to address AIOps for failure management tasks. As shown in Figure~\ref{fig: llm-approach}, in this survey, we categorize these approaches into four groups: foundation model, fine-tuning approach, embedding-based approach, and prompt-based approach. The foundation model refers to pre-trained language models that serve as the base for AIOps tasks without further modifications. The fine-tuning approach involves adapting these pre-trained models to specific tasks through additional training on task-specific data. The embedding-based approach uses the representations generated by pre-trained models to capture semantic information and improve task performance. The prompt-based approach leverages natural language prompts to guide the model's responses, enabling it to perform specific tasks based on the given instructions.

\textbf{Foundational Model.} In the foundational models pre-trained for AIOps tasks, most are based on the Transformer framework~\cite{liao2024timegpt, garza2023timegpt, rasul2023lag, das2023decoder, gupta2023learning, shi2023shellgpt, dong2024simmtm, mastropaolo2022using}, though a few are not~\cite{ekambaram2023tsmixer, ekambaram2024ttms, wu2022timesnet, wang2023observed}.

\begin{itemize}
	\item \textbf{Transformer-based.} These models are built on the Transformer architecture, which uses self-attention mechanisms to capture dependencies in data. Examples include BERT, GPT, and T5, which have been widely used for various NLP tasks due to their ability to understand and generate human language effectively.
	\item \textbf{non-Transformer-based.} This category includes models such as MLP (Multi-Layer Perceptron), RNN (Recurrent Neural Networks), CNN (Convolutional Neural Networks), diffusion-based models, and others. These models leverage different architectural principles and have unique strengths in handling specific types of data and tasks. For instance, RNNs are particularly suited for sequential data, while CNNs excel in processing grid-like data structures.
\end{itemize}

\textbf{Fine-tuning Approach.} Directly applying general foundational models to AIOps for failure management tasks often does not yield optimal results. Therefore, many AIOps approaches fine-tune these models using domain-specific datasets~\cite{sarda2023leveraging, bhavya2023exploring, guo2023owl, zhou2023one, jin2023time, chang2023llm4ts, ahmed2023recommending, chen2022bert, dang2021ts}. This fine-tuning can be categorized into two types: \textbf{full fine-tuning} and \textbf{parameter-efficient fine-tuning}.

\begin{itemize}
	\item \textbf{Full Fine-Tuning.} This involves updating all the parameters of the foundational model using domain-specific data. However, due to the large number of parameters in models like GPT-3.5 and GPT-4, most current AIOps work in this area is based on models like BERT and T5, which are more manageable in size and complexity.
	\item \textbf{Parameter-Efficient Fine-Tuning.} This includes techniques such as Layer-Freezing, Adapter Tuning, and Task-Conditional Fine-Tuning. These methods require tuning only a small subset of the model's parameters, making them more efficient and practical for AIOps applications. Layer-Freezing involves freezing most of the model's layers and only fine-tuning the top layers, allowing for significant reductions in computational cost and training time. Adapter Tuning introduces small, trainable adapter modules within each layer of the pre-trained model, enabling the model to adapt to new tasks with minimal effort. Task-Conditional Fine-Tuning involves adding task-specific output heads to the model's output and training them accordingly. Parameter-efficient fine-tuning is the predominant approach in the AIOps field.
\end{itemize}

\textbf{Embedding-based Approach.} Many data sources for AIOps contain rich semantic information, such as logs, documentation, etc., and embeddings play a crucial role in reflecting this semantic information in a data format. Therefore, many AIOps works are based on embeddings. These embedding-based approaches can be categorized into two types: \textbf{pre-trained embedding}~\cite{ott2021robust, shao2022log, karlsen2023exploring, hu2023research, almodovar2024logfit, zhang2022logst} and \textbf{prompt embedding}~\cite{sun2023test, gruver2024large, cao2023tempo}. Pre-trained embedding works involve using LLMs to embed this information to capture its semantic information, while prompt embedding works design specific embedding methods suitable for LLMs to activate the LLM's processing capability for specific data.

\begin{itemize}
	\item \textbf{Pre-Trained Embedding.} Pre-trained embedding works involve directly leveraging embeddings generated by LLMs, such as BERT, GPT, or T5, to capture the semantic information of various data sources in AIOps. These embeddings are obtained from models pre-trained on large-scale text corpora and capture rich semantic information, making them suitable for a wide range of AIOps tasks without further fine-tuning. Currently, most works related to logs utilize this embedding-based approach.
	\item \textbf{Prompt Embedding.} Prompt embedding works focus on designing specific embedding methods tailored for LLMs to effectively capture semantic information from various data sources in AIOps. By providing natural language prompts or instructions, these methods activate the LLM's processing capability to generate task-specific embeddings. This approach enables flexible and task-specific embeddings suitable for different AIOps applications. Currently, many works based on metrics utilize this approach to transform metrics data into a format more suitable for LLM understanding.
\end{itemize}

\textbf{Prompt-based Approach.} In addition to using the embedding capabilities of large language models (LLMs), many AIOps works directly input prompts into LLMs to complete failure management tasks~\cite{liu2023opseval, park2023formulating, cao4741492managing, xue2023promptcast, liu2024lstprompt, liu2023scalable, zhang2023logprompt, li2022evaluating, quan2023heterogeneous, aiello2023service}. These works can be categorized into \textbf{In-Context Learning (ICL)}, \textbf{Chain of Thoughts (CoT)}, \textbf{Task Instruction Prompting}, and \textbf{Knowledge-Based Approach}.

\begin{itemize}
	\item \textbf{In-Context Learning (ICL).}  In-Context Learning involves providing the model with examples within the input prompt to help it understand how to perform a task. This approach allows the model to generate appropriate responses based on the patterns and information presented in the provided context. Many AIOps works use this method to guide LLMs in producing results that conform to the desired output format.
	\item \textbf{Chain of Thoughts (CoT).}  Chain of Thoughts involves guiding the model through a logical sequence of intermediate steps or reasoning processes to arrive at the final answer. This approach helps in complex problem-solving tasks where the solution requires multi-step reasoning. Many AIOps works use this method to improve the accuracy of the output.
	\item \textbf{Task Instruction Prompting.} Task Instruction Prompting entails giving explicit instructions to the model about the task to be performed. This method leverages the model's ability to follow detailed natural language instructions to complete specific tasks effectively. In this survey, we categorize many simple or zero-shot methods under this approach. Many early AIOps works use this method to accomplish specific tasks.
	\item \textbf{Knowledge-Based Approach.} This approach integrates external knowledge into the model's responses. Examples include Tool Augmented Generation (TAG), where the model uses external tools or APIs to enhance its capabilities, and Retrieval Augmented Generation (RAG), which involves retrieving relevant information from external sources to improve the model's responses. This is the most commonly used method in AIOps works and is the most effective in improving LLM performance for specific tasks.
\end{itemize}

\subsection{AIOps Tasks for Failure Management}

\begin{figure}[htbp]
	\centering
	\tikzset{
		my node/.style={
			draw,
			align=center,
			thin,
			text width=1.2cm, 
			rounded corners=3,
		},
		my leaf/.style={
			draw,
			align=left,
			thin,
			text width=8.5cm, 
			rounded corners=3,
		}
	}
	\forestset{
		every leaf node/.style={
			if n children=0{#1}{}
		},
		every tree node/.style={
			if n children=0{minimum width=1em}{#1}
		},
	}
	\begin{forest}
		nonleaf/.style={font=\bfseries\scriptsize},
		for tree={%
			every leaf node={my leaf, font=\scriptsize},
			every tree node={my node, font=\scriptsize, l sep-=4.5pt, l-=1.pt},
			anchor=west,
			inner sep=2pt,
			l sep=10pt, 
			s sep=3pt, 
			fit=tight,
			grow'=east,
			edge={ultra thin},
			parent anchor=east,
			child anchor=west,
			if n children=0{}{nonleaf}, 
			edge path={
				\noexpand\path [draw, \forestoption{edge}] (!u.parent anchor) -- +(5pt,0) |- (.child anchor)\forestoption{edge label};
			},
			if={isodd(n_children())}{
				for children={
					if={equal(n,(n_children("!u")+1)/2)}{calign with current}{}
				}
			}{}
		}
		[LLM-based AIOps Tasks \\for Failure Management, draw=gray, fill=gray!15, text width=2.8cm, text=black
		[Data Preprocessing \\(\color{blue}{\cref{sec:data_preprocess}}\color{black}), color=brightlavender, fill=brightlavender!15, text width=3cm, text=black
		[Log Parsing (\color{blue}\cref{sec:log_parsing}\color{black}), color=brightlavender, fill=brightlavender!15, text width=5cm, text=black],
		[Metrics Imputation (\color{blue}\cref{sec:metrics_imputation}\color{black}), color=brightlavender, fill=brightlavender!15, text width=5cm, text=black],
		[Input Summarization (\color{blue}\cref{sec:input_summarization}\color{black}), color=brightlavender, fill=brightlavender!15, text width=5cm, text=black]
		]
		[Failure Perception \\ (\color{blue}\cref{sec:failure_perception}\color{black}), color=lightgreen, fill=lightgreen!15, text width=3cm, text=black
		[Failure Prediction  (\color{blue}\cref{sec:failure_prediction}\color{black}), color=lightgreen, fill=lightgreen!15, text width=5cm, text=black]
		[Anomaly Detection  (\color{blue}\cref{sec:anomaly_detection}\color{black}), color=lightgreen, fill=lightgreen!15, text width=5cm, text=black]
		]
		[Root Cause Analysis \\ (\color{blue}\cref{sec:root_cause_analysis}\color{black}), color=harvestgold, fill=harvestgold!15, text width=3cm, text=black
		[Failure Localization (\color{blue}\cref{sec:failure_localization}\color{black}), color=harvestgold, fill=harvestgold!15, text width=5cm, text=black]
		[Failure Category Classification  (\color{blue}\cref{sec:failure_category_classification}\color{black}), color=harvestgold, fill=harvestgold!15, text width=5cm, text=black],
		[Root Cause Report Generation (\color{blue}\cref{sec:root_cause_report_generation}\color{black}), color=harvestgold, fill=harvestgold!15, text width=5cm, text=black]
		]
		[Auto Remediation \\ (\color{blue}\cref{sec:auto_remediation}\color{black}), color=carminepink, fill=carminepink!15, text width=3cm, text=black
		[Assisted Questioning (\color{blue}\cref{sec:assisted_questioning}\color{black}), color=carminepink, fill=carminepink!15, text width=5cm, text=black]
		[Mitigation Solution Generation (\color{blue}\cref{sec:mitigation_solution_generation}\color{black}), color=carminepink, fill=carminepink!15, text width=5cm, text=black],
		[Command Recommendation (\color{blue}\cref{sec:script_recommendation}\color{black}), color=carminepink, fill=carminepink!15, text width=5cm, text=black]
		[Script Generation (\color{blue}\cref{sec:script_generation}\color{black}), color=carminepink, fill=carminepink!15, text width=5cm, text=black]
		[Automatic Execution (\color{blue}\cref{sec:automatic_execution}\color{black}), color=carminepink, fill=carminepink!15, text width=5cm, text=black]
		]
		]
	\end{forest}
	\caption{AIOps Tasks for Failure Management (taxonomy of this survey)}
	\label{fig: taxonomy}
\end{figure}
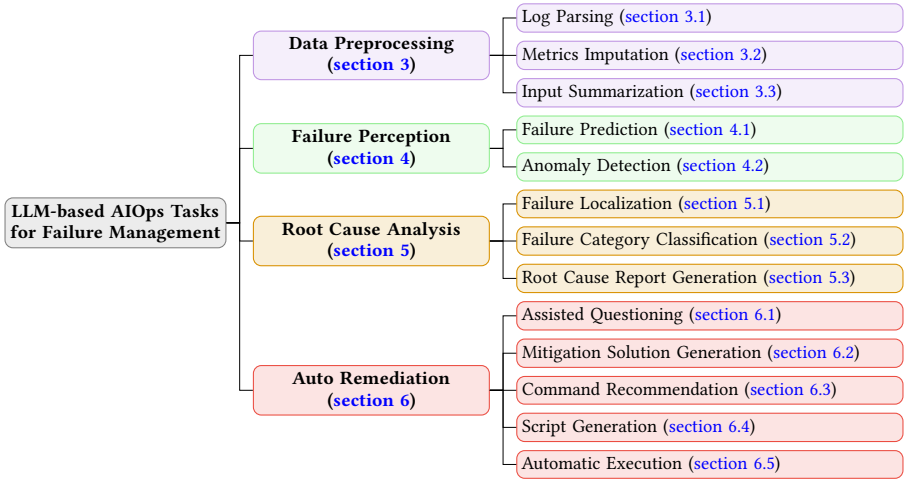

AIOps tasks for failure management comprise a holistic goal consisting of several subtasks. As shown in Figure~\ref{fig: taxonomy}, the entire process can be divided into data preprocessing, failure perception, root cause analysis, and auto remediation. These tasks are sequentially related: failure perception is based on preprocessed data, and once an anomaly is detected in the software system, root cause analysis is conducted to automatically identify where and what the anomaly is. Finally, after determining the cause of the anomaly, appropriate auto remediation methods are implemented to mitigate the software system issue.

\textbf{Data Preprocessing.} In the context of AIOps tasks, data preprocessing does not refer to conventional processes like data scraping or cleaning. Instead, it involves tasks specific to the AIOps domain. With the integration of LLMs, these preprocessing tasks can be categorized into three types: \textbf{Log Parsing}~\cite{mudgal2023assessment, guo2023owl, jiang2023llmparser, ma2024llmparser, guo2024lemur, le2023log, gupta2023learning, xu2023prompting, sun2023design}, \textbf{Metrics Imputation}~\cite{zhou2023one, gruver2024large, chen2023gatgpt, wang2023observed}, and \textbf{Input Summarization}~\cite{mudgal2023assessment, zhang2024automated, goel2024x, egersdoerfer2023early, chen2024automatic, zhou2023d, kuang2024knowledge, jin2023assess}.

\begin{itemize}
	\item \textbf{Log Parsing.} Log parsing involves analyzing and structuring log data to extract meaningful information. This task typically includes identifying patterns, categorizing log entries, and converting unstructured log data into structured formats that are easier to analyze. For example, the log template "E0,('instruction', 'cache', 'parity', 'error', 'corrected')" can be extracted from the log message “2005-06-03-15.42.50.363779 R02-M1-N0-C:J12-U11 RAS KERNEL INFO instruction cache parity error corrected". There are already many log parsing methods, based on frequent pattern mining~\cite{dai2020logram, vaarandi2015logcluster, sedki2022effective, yu2023brain}, clustering~\cite{hamooni2016logmine, shima2016length, tang2011logsig, nedelkoski2021self}, and heuristics~\cite{he2017drain, jiang2008abstracting, makanju2009clustering}. However, LLMs can enhance log parsing by using natural language understanding to accurately interpret and classify log data.
	\item \textbf{Metrics Imputation.} Metrics imputation deals with estimating missing or incomplete metrics data. This is crucial for maintaining accurate and comprehensive datasets for failure perception and other analysis. LLMs can improve metrics imputation by leveraging contextual information and patterns in the available data to make accurate estimations of missing values.
	\item \textbf{Input Summarization.} Input summarization focuses on condensing extensive text into concise summaries that highlight the most important information. This task helps in quickly identifying significant events or issues within the input text. Since many large models have limitations on the number of tokens that can be processed in a single query, and log data can be vast, input summarization often serves as a preliminary step for many log-based tasks. This step is typically performed using LLMs, as they can generate effective summaries by understanding the context and extracting key points from large volumes of text data.
\end{itemize}

\textbf{Failure Perception.} After completing data preprocessing, failure perception is the next crucial step in identifying deviations from normal behavior in a system. This process is essential for early detection of potential issues, enabling proactive measures to prevent failures. These approaches can be categorized into two types: \textbf{Failure Prediction}~\cite{xiong2023can, alharthi2022clairvoyant, alharthi2023time, yang2023diffusion} and \textbf{Anomaly Detection}~\cite{zhou2023one, xue2023promptcast, jin2023time, chang2023llm4ts, gruver2024large, cao2023tempo, garza2023timegpt, rasul2023lag, das2023decoder, liu2024lstprompt, li2022evaluating, dang2021ts, karlsen2024large, alnegheimish2024large, liu2024large, sun2023test, liu2023scalable, karlsen2023exploring, dong2024simmtm, almodovar2024logfit, zhang2022logst, karlsen2024large, mudgal2023assessment, pan2023raglog, guo2023owl, egersdoerfer2023early, qi2023loggpt, chen2022bert, lee2023lanobert, ott2021robust, zhang2023logprompt, huang2023improving, shao2022log, he2023parameter, hu2023research, le2021log, huang2020hitanomaly, sun2023design}.

\begin{itemize}
	\item \textbf{Failure Prediction.} Failure prediction involves forecasting potential system failures before they occur by analyzing historical data and identifying patterns that precede failures. This task enables proactive measures by alerting maintenance personnel to potential issues in advance, allowing for early remediation. However, there has been relatively limited work on failure prediction in the era of LLMs, as this approach often faces challenges related to data quality and the complexity of accurately predicting failures in dynamic environments.
	\item \textbf{Anomaly Detection.} Due to the limitations of failure prediction, most work has shifted towards anomaly detection. Anomaly detection aims to identify abnormal behavior or patterns that deviate from the norm, indicating potential issues or failures. This approach is more adaptable to the dynamic nature of software systems and is widely used in LLM-based methods. LLMs can effectively analyze large volumes of data, including logs and metrics, to detect subtle anomalies that traditional methods might miss. This capability is crucial for maintaining the reliability and availability of large-scale distributed systems, as it allows for the timely identification and resolution of issues before they escalate into critical failures.
\end{itemize}

\textbf{Root Cause Analysis.} After detecting anomalies in a software system, it is crucial to analyze which component is experiencing issues and what specific anomaly is occurring. Accurate root cause analysis can greatly assist operations personnel in the subsequent repair process. These approaches, depending on the task performed, can be categorized into three types: \textbf{Failure Localization}~\cite{li2024llm, sarda2023adarma, sarda2023leveraging, roy2024exploring, shan2024face}, \textbf{Failure Category Classification}~\cite{chen2024automatic, zhang2023pace, zhou2023one, sun2023test, gupta2023learning, zhou2024llm, zhou2023d, quan2023heterogeneous}, \textbf{Root Cause Report Generation}~\cite{roy2024exploring, zhang2024automated, wang2023rcagent, goel2024x, ahmed2023recommending}.

\begin{itemize}
	\item \textbf{Failure Localization.} This task aims to identify the specific component or machine where the anomaly occurred, often using methods such as Causal Discovery. In microservices environments, it can pinpoint the exact service or machine experiencing the issue. Additionally, it may involve identifying the specific log or metric entry that marked the onset of the anomaly. While widely used in traditional, non-LLM-based works, this type of work is less common in current LLM-based approaches.
	\item \textbf{Failure Category Classification.} This task identifies what type of anomaly the system is experiencing, such as CPU resource shortages, memory shortages, or software configuration errors. The advent of LLMs has enhanced the cross-platform generality of these methods and expanded their ability to classify a wider range of anomalies. Currently, this task typically uses incident reports as input.
	\item \textbf{Root Cause Report Generation.} This task directly generates a root cause analysis report, leveraging LLMs' powerful natural language generation capabilities. It generally combines the previous two tasks, providing a comprehensive report that details the components and reasons behind the software system's anomalies.
\end{itemize}

\textbf{Auto Remediation.} After identifying the type and location of the software system anomaly, the next step is to automatically mitigate and repair the issue based on this information. Prior to the advent of LLMs, the automation level of these remediation tasks was relatively low. However, with the introduction of LLMs, the degree of automation has seen a significant improvement. Auto remediation tasks, in increasing level of automation, can be categorized as follows: \textbf{Assisted Questioning}~\cite{liu2023opseval, park2023formulating, guo2023owl}, \textbf{Mitigation Solution Generation}~\cite{goel2024x, wang2023network, ahmed2023recommending, hamadanian2023holistic}, \textbf{Command Recommendation}~\cite{xue2023db, shi2023shellgpt}, \textbf{Script Generation}~\cite{sarda2023leveraging, jiang2024xpert, wang2023low, pesl2023uncovering, aiello2023service}, \textbf{Automatic Execution}~\cite{cao4741492managing, khlaisamniang2023generative, othman2023fostering, sarda2023adarma}.

\begin{itemize}
	\item \textbf{Assisted Questioning.} This task involves using LLMs to assist operations personnel by answering system-related questions. By allowing operations staff to directly query the LLM software, detailed responses can be obtained quickly, speeding up the resolution of software system failures. This type of task emerged with the rise of LLMs, particularly GPT-3.5.
	\item \textbf{Mitigation Solution Generation.} In this task, LLMs generate potential mitigation solutions for the detected anomalies. These solutions are often derived from large datasets of historical incident reports and resolutions, providing actionable suggestions to operations personnel for addressing issues. Before LLMs, auto remediation for software failures primarily involved incident triage~\cite{shao2008efficient, zeng2017knowledge}, but these methods are not intelligent. Mitigation solution generation appears more advanced and tailored.
	\item \textbf{Command Recommendation.} This task leverages LLMs to recommend pre-existing scripts that can be used to remediate the detected issue. The system can suggest the next command to enter when operations staff input commands (e.g., shell commands), thus shortening the repair time. This method gaines popularity with the advent of LLMs.
	\item \textbf{Script Generation.} This task skips the command recommendation step, directly using LLMs to generate custom scripts tailored to resolve specific detected anomalies. This involves creating new scripts based on the details of the issue and the context provided by system logs and metrics, allowing for more precise and effective remediation actions.
	\item \textbf{Automatic Execution.} The highest level of automation involves LLMs not only generating the necessary remediation scripts but also executing them automatically. This end-to-end process ensures that detected anomalies are addressed without manual intervention, significantly speeding up resolution time and reducing the workload on operations personnel. Although this method is highly attractive, related work is limited, and its practical effectiveness remains to be verified.
\end{itemize}

The survey will follow the structure outlined in Figure~\ref{fig: taxonomy} to provide a detailed overview of LLM-based AIOps tasks for failure management. Section~\blueref{sec:data_preprocess} will delve into the related work on data preprocessing. Section~\blueref{sec:failure_perception} will comprehensively discuss the work on failure perception. Section~\blueref{sec:root_cause_analysis} will cover the efforts related to root cause analysis. Finally, Section~\blueref{sec:auto_remediation} will explore the work on auto remediation.

\section{Data Preprocessing}
\label{sec:data_preprocess}

As aforementioned, the data sources for AIOps can be categorized into two groups: system-generated data and human-generated data. Human-generated data, which come in various forms, are typically processed using natural language methods without special handling. In contrast, system-generated data are usually of fixed types and relatively uniform formats, but they tend to be larger in volume and continuously growing. Therefore, most data preprocessing methods are designed to handle system-generated data. In this survey, we categorize these preprocessing methods into three types: log parsing (Section~\blueref{sec:log_parsing}), input summarization (Section~\blueref{sec:input_summarization}), and metrics imputation (Section~\blueref{sec:metrics_imputation}).

\subsection{Log Parsing}
\label{sec:log_parsing}

Log parsing is fundamental to many tasks that use logs as inputs for failure perception and root cause analysis, serving as the basis for subsequent tasks.

\begin{figure}[h]
	\centering
	\includegraphics[width=\textwidth]{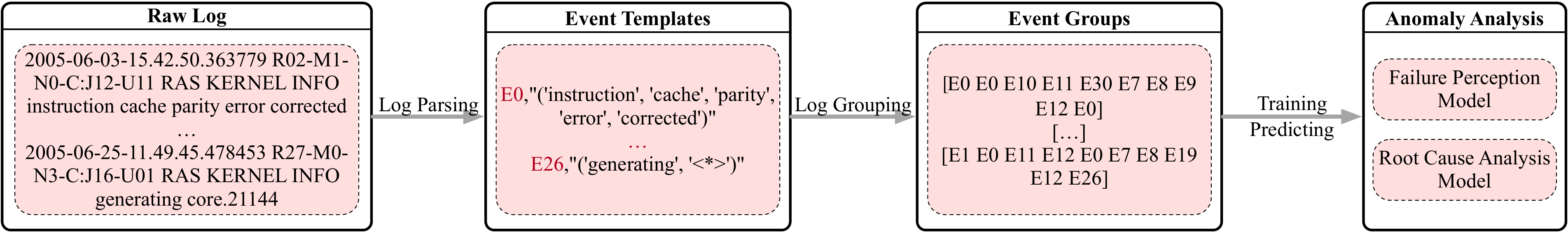}
	\caption{Log-based Failure Perception and Root Cause Analysis: The Common Workflow}
	\label{fig: common-workflow}
\end{figure}

As shown in Figure~\ref{fig: common-workflow}, raw logs consist of semi-structured text encompassing various fields like timestamps and severity levels. For the benefit of downstream tasks, log parsing is employed to transform each log message into a distinct event template, which includes a constant part paired with variable parameters. For example, the log template "E0,('instruction', 'cache', 'parity', 'error', 'corrected')" can be extracted from the log message “2005-06-03-15.42.50.363779 R02-M1-N0-C:J12-U11 RAS KERNEL INFO instruction cache parity error corrected" in Figure~\ref{fig: common-workflow}. After being parsed into event templates, log data can be organized into sequence groups using session, sliding, or fixed windows. Following this, failure perception and root cause analysis are performed on each event group to determine if a failure exists, and if so, to conduct the corresponding root cause analysis.

Next, we provide a formal description of this process. For a given log sequence $S=(s_{1}, s_{2}, ..., s_{N})$, where $s_{n}$ represents an individual log entry. After log parsing, each log $s_{n}$ can be represented as an event $e_{i}$, and the entire collection of unique events can be denoted as $\omega = \{e_{1} , e_{2} , ..., e_{n}\}$.  After log grouping, the entirety of raw log messages $S_{t}=(s_{1}, s_{2}, ..., s_{M_{t}})$ in a specific time window $t$ can be represented as $E_{t} = (e_{(s_{1})}, e_{(s_{2})}, ..., e_{(s_{M^{t}})})$. Here, $M^{t}$ represents the length of each grouped log sequence in the time window $t$.

A common issue in traditional log parsing methods is their lack of generalizability. These methods often rely on manually designed rules or are trained on limited datasets. As a result, their effectiveness significantly decreases when applied to different software systems or when there are changes in log generation rules. The emergence of powerful LLMs, which possess extensive pre-trained knowledge related to code and logging, offers a promising avenue for log parsing. However, the lack of specialized log parsing capabilities currently hinders the accuracy of LLMs in this task. Additionally, the inherent inconsistencies in their responses and the substantial computational overhead prevent the practical adoption of LLM-based log parsing. Consequently, many studies have explored the effectiveness of LLMs for log parsing and have proposed methods to effectively leverage LLMs in this domain.

\textbf{Empirical Study.} Priyanka et al.~\cite{mudgal2023assessment} conducted a study on zero-shot log parsing using ChatGPT. Their findings indicated that the current version of ChatGPT has limited performance in zero-shot log processing, with issues of response inconsistency and scalability. Le et al.~\cite{le2023log} evaluated ChatGPT's ability to perform log parsing through two research questions: the effectiveness of ChatGPT in parsing logs and its performance with different prompting methods. Their results showed that ChatGPT can achieve promising outcomes in log parsing with appropriate prompts, especially with a few-shot approach. These empirical studies demonstrate that while LLMs have the potential for log parsing, they require effective methods to guide them.

\textbf{Prompt-based Approach.} Some studies have adopted prompt-based methods to guide LLMs for effective log parsing. LILAC~\cite{jiang2023llmparser} proposes a practical log parsing framework using LLMs with an adaptive parsing cache. LILAC leverages the in-context learning (ICL) capabilities of LLMs by executing a hierarchical candidate sampling algorithm and selecting high-quality demonstrations. It uses an adaptive parsing cache to store and optimize LLM-generated templates, helping mitigate the inefficiency of LLMs by quickly retrieving previously processed log templates. LLMParser~\cite{ma2024llmparser} employs in-context learning and few-shot tuning methods. This approach is tested on four LLMs: Flan-T5-small, Flan-T5-base, LLaMA-7B, and ChatGLM-6B, across 16 open-source systems, and find that smaller LLMs might be more effective for log parsing tasks than more complex ones. Lemur~\cite{guo2024lemur} introduces a novel sampling method inspired by information entropy, effectively clustering typical logs and using the Chain of Thoughts (CoT) method with LLMs to distinguish parameters from invariant tokens. DivLog~\cite{xu2023prompting} combines Retrieval-Augmented Generation (RAG) and ICL methods for log parsing, sampling a diverse set of offline logs as candidate logs and selecting five suitable template candidates for each target log during the parsing process. Sun et al.~\cite{sun2023design} develop a cloud-native log management platform providing log collection, transmission, storage, and system management features, utilizing GPT-3.5 for few-shot log parsing.

\textbf{Fine-tuning Approach.} Other works have employed fine-tuning methods to train pre-trained models specifically for log parsing. OWL~\cite{guo2023owl} utilizes supervised fine-tuning and mixture adapter tuning methods to train a set of large language models for knowledge querying and log parsing based on the LLaMA model and their OWL-Instruct dataset. Pranjal et al.\cite{gupta2023learning} applies full fine-tuning on BERT with 12 Loghub datasets\cite{he2023loghub} and 5 proprietary data sources, creating a specialized LLM named BERTOps that effectively performs multiple downstream log tasks, including log parsing.

\subsection{Metrics Imputation}
\label{sec:metrics_imputation}

In metrics-based methods, a significant issue arises due to the sheer volume of metrics data—potentially thousands of indicators generating data points every second. During transmission or recording, data loss can easily occur. These losses might not necessarily indicate anomalies in the software system at the time. Therefore, it's crucial to impute these missing metrics data points, which can effectively enhance the performance of subsequent tasks.

The metrics imputation process can be formalized as follows. Given a time series of metrics data $M = \{M_1, M_2, ..., M_T\}$ over time $T$, where each metric $M_t$ consists of a set of measurements $\{m_t^1, m_t^2, ... , m_t^n\}$ for $n$ different indicators, metrics imputation involves filling in the missing values in this data.

Let $M^{'} = \{M_1^{'}, M_2^{'}, ..., M_T^{'}\}$ be the observed metrics data with missing values. For each time step $t$, $M_t^{'}$ includes both observed measurements and missing measurements. The goal is to estimate the missing values $M_t^{'}$ such that the imputed data $M^{*} = \{M_1^{*}, M_2^{*}, ..., M_T^{*}\}$ approximates the original complete data $M$.

Formally, assume $M_t = \{m_t^1, m_t^2, ... , m_t^n\}$ and let $O_t \subseteq {1,2,...,n}$ be the set of indices of observed values at time $t$. Then the imputed metrics at time $t$ can be represented as Equation~\ref{eq: metrics-imputation-input}, where \(\hat{m}_t^i\) is the imputed value for the missing metric at index \(i\) and time \(t\).

\begin{equation}
	\left\{
	\begin{array}{ll}
		m_t^i & \text{if } i \in \mathcal{O}_t \\
		\hat{m}_t^i & \text{if } i \notin \mathcal{O}_t
	\end{array}
	\right.
	\label{eq: metrics-imputation-input}
\end{equation}

The imputation process aims to minimize the difference between the true values \(m_t^i\) and the imputed values \(\hat{m}_t^i\) for all missing entries. This can be formalized as an optimization problem as Equation~\ref{eq: metrics-imputation}. Thus, the task of metrics imputation involves accurately estimating \(\hat{m}_t^i\) to reconstruct the complete metrics dataset $M$ from the observed dataset $M^{'}$.

\begin{equation}
	\min_{M^{*}} \sum_{t=1}^{T} \sum_{i \notin \mathcal{O}_t} \left( m_t^i - \hat{m}_t^i \right)^2
	\label{eq: metrics-imputation}
\end{equation}

The metrics imputation approaches can be categorized into two types based on their ability to yield varied imputations that reflect the inherent uncertainty in the imputation process~\cite{wang2024deep}: predictive methods and generative methods.

\textbf{Predictive Approach.} Predictive imputation methods consistently predict deterministic values for the same missing components. Many works have proposed deep learning models for this purpose. For instance, GRU-D~\cite{che2018recurrent} employs RNN models, TimesNet~\cite{wu2022timesnet} utilizes CNN models, and Saits~\cite{du2023saits} is based on attention models. Additionally, several approaches leverage large language models (LLMs). Zhou et al.~\cite{zhou2023one} fine-tuned pre-trained models like GPT-2 and BERT using layer-freezing methods, while Nate et al.~\cite{gruver2024large} used prompt embedding techniques with GPT-3, LLaMA-2, and GPT-4 to accomplish metrics imputation tasks. GatGPT~\cite{chen2023gatgpt} employs a graph attention network to pre-train a large language model specifically designed for metrics imputation.

\textbf{Generative Approach.} Generative methods are built upon models like VAEs, GANs, and diffusion models. There are fewer works in this category that are based on LLMs. GP-VAE~\cite{fortuin2020gp}, V-RIN~\cite{mulyadi2021uncertainty}, and supnotMIWAE~\cite{kim2023probabilistic} use VAE methods for metrics imputation, employing an encoder-decoder structure to approximate the true data distribution by maximizing the Evidence Lower Bound (ELBO) on the marginal likelihood. NAOMI~\cite{liu2019naomi} and USGAN~\cite{miao2021generative} utilize GANs for generative metrics imputation. SSSD~\cite{alcaraz2022diffusion} and CSBI~\cite{chen2023provably} are based on diffusion models, which capture complex data distributions by progressively adding and then reversing noise through a Markov chain of diffusion steps.

\subsection{Input Summarization}
\label{sec:input_summarization}

Since the context space of LLMs is always limited, providing extensive data as external knowledge or associative information requires summarizing this data first. The summarized results can then be merged into a single context for submission to the LLM.

Input summarization is essentially an information compression process. It takes long inputs and summarizes them so that the summarized result contains as much effective information as the original input while significantly reducing the input length.

This process can be formalized as follows. For a given original input sequence $S=(s_{1}, s_{2}, ..., s_{N})$, where $s_{n}$ represents represents a segment of the input. The goal is to produce a summarized sequence $\hat S$, which can be represented as $\hat S=(\hat s_{1}, \hat s_{2}, ..., \hat s_{M})$, where $M < N$ and $\hat S$ retains the essential information from $S$. The summarization function $f$ can be defined as Equation~\ref{eq: input-summarization}, where $f$ aims to minimize the loss of information, ensuring that $\hat S$ is a compact representation of $S$ with maximum information retention.

\begin{equation}
	\hat S = f(S)
	\label{eq: input-summarization}
\end{equation}

This summarized sequence $\hat S$ is then used as the input to the LLM, effectively providing the model with a condensed version of the original data, allowing it to process and utilize the information within the constraints of its context window. We categorize the methods into two types based on whether an LLM is used as the compression tool during the summarization process.

\textbf{non-LLM-based Approach.} This type of approach does not use LLMs as compression tools~\cite{zhou2024survey}. In AIOps for failure management, these methods generally follow three main strategies. DYNAICL~\cite{zhou2023efficient}, Selective Context~\cite{li2023compressing}, LLMLingua~\cite{jiang2023llmlingua}, and LongLLMLingua~\cite{jiang2023longllmlingua} employ prompt pruning to remove unimportant tokens, sentences, or documents from each input prompt online based on predefined or learnable importance indicators. RECOMP~\cite{xu2023recomp} and SemanticCompression~\cite{fei2023extending} condense the original prompt into a shorter summary while preserving similar semantic information. Many metric-based methods, such as Nate et al.~\cite{gruver2024large} and TEST~\cite{sun2023test}, use soft prompt-based strategies to design a special prompt tailored to specific scenarios and data, which is significantly shorter than the original prompt for use as input to LLMs.

\textbf{LLM-based Approach.} This type of approach directly uses LLMs as compression tools. Priyanka et al.~\cite{mudgal2023assessment} and Chris et al.~\cite{egersdoerfer2023early} employ zero-shot methods by directly inputting prompts to test ChatGPT's compression effectiveness on log data. Zhang et al.~\cite{zhang2024automated}, Drishti et al.~\cite{goel2024x}, Chen et al.~\cite{chen2024automatic}, and Oasis~\cite{jin2023assess} attempt to summarize large volumes of incident reports. Zhou et al.~\cite{zhou2024llm} and D-Bot~\cite{zhou2023d} summarize external technical documents or code to create auxiliary knowledge bases.

\section{Failure Perception}
\label{sec:failure_perception}

Failure perception is conducted based on the preprocessed data and serves as the most critical step in AIOps for failure management. It is used to predict or detect whether anomalies have occurred in the runtime software system, forming the foundation for subsequent root cause analysis and auto remediation. Typically, this step involves the continuous online perception using system-generated data. In this survey, we categorize the failure perception into two subtasks: failure prediction (Section~\blueref{sec:failure_prediction}) and anomaly detection (Section~\blueref{sec:anomaly_detection}).

\subsection{Failure Prediction}
\label{sec:failure_prediction}

Failure prediction involves analyzing historical data to detect whether a software system is likely to experience a failure within a future time window. This process helps system operators proactively identify and address issues before they lead to actual failures. 

Formally, given a time series of historical system data $X = \{x_1, x_2, \ldots, x_t\}$, the goal is to predict the probability $P(f_{t+\Delta t})$ of a failure $f$ occurring within a future interval $\Delta t$. This process can be expressed as $P(f_{t+\Delta t} | X)$, which is the predicted probability of a failure occurring at time $t + \Delta t$ given the historical data $X$.

Although failure prediction offers invaluable foresight, empowering operations teams to proactively address impending issues, the realm of LLM-based approaches in this domain remains sparse. This scarcity stems from the inherent challenge that numerous failures evade advanced notice, resulting in methodologies that are either confined to addressing a narrow spectrum of anomalies or suffer from high rates of missed detections (false negatives).

Nonetheless, certain research endeavors harness LLMs to augment the efficacy of failure prediction models, albeit in a supplementary capacity. Clairvoyant~\cite{alharthi2022clairvoyant} leverages log data to train a novel self-supervised model based on BERT to predict node failures in HPC systems. Clairvoyant can predict only failures, while Time Machine~\cite{alharthi2023time} builds upon it with a two-stack transformer-decoder architecture to predict not only failures but also their lead times. Yang et al.~\cite{yang2023diffusion} focus on enhancing data quality through data imputation to improve the performance of the downstream failure prediction task based on a sample-efficient diffusion model. Xiong et al.~\cite{xiong2023can} define a systematic framework based on a prompting approach with three components: prompting strategies for eliciting verbalized confidence, sampling methods for generating multiple responses, and aggregation techniques for computing consistency. They benchmark these methods on the failure prediction task using five widely-used LLMs, including GPT-4 and LLaMA2 Chat, and uncover that LLMs tend to be overconfident, potentially imitating human patterns of expressing confidence. However, employing their proposed strategies, such as human-inspired prompts, consistency among multiple responses, and better aggregation strategies, can help mitigate this overconfidence from various perspectives and achieve results close to those of fine-tuned models.

\subsection{Anomaly Detection}
\label{sec:anomaly_detection}

\begin{figure}[h]
	\centering
	\includegraphics[width=\textwidth]{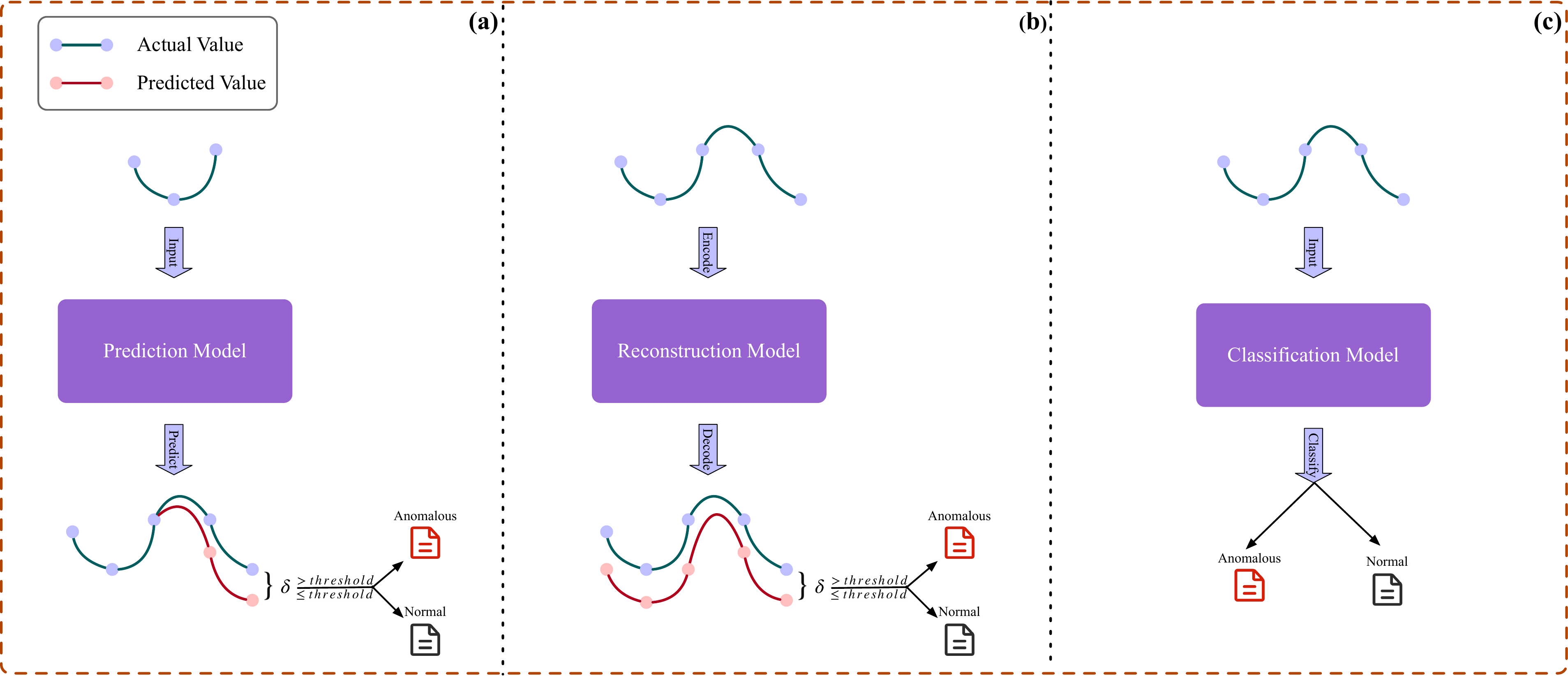}
	\caption{Workflow of Various Anomaly Detection Approaches: Prediction-based, Reconstruction-based, and Classification-based}
	\label{fig: anomaly-detection}
\end{figure}

Unlike failure prediction, anomaly detection involves analyzing historical data to determine whether anomalies exist within a specific time window. Based on the methods used for detecting anomalies, as illustrated in Figure~\ref{fig: anomaly-detection}, anomaly detection approaches can be categorized into three types: \textbf{Prediction-Based Methods}, \textbf{Reconstruction-Based Methods}, and \textbf{Classification-Based Methods}.

\textbf{Prediction-Based Methods.} Prediction-based anomaly detection forecast future system data based on historical data and identify anomalies by comparing the predicted values with the actual observed values. Significant deviations between the prediction and the actual values are flagged as anomalies. This process can be viewed as a time-series forecasting task for metrics data and an event prediction task for logs data. However, regardless of the data type, the general process can be uniformly formalized.

Formally, assume $X = \{x_1, x_2, ..., x_T\}$ represent the historical data, where $x_t$ is the observed value at time $t$. The goal is to predict the future values $\hat{X} = \{\hat{x}_{T+1}, \hat{x}_{T+2}, ..., \hat{x}_{T+\tau}\}$, where $\tau$ is the prediction horizon. The prediction process of each data point can be represented as Equation~\ref{eq: prediction}, where $k = 1, 2, ..., \tau$, and $f$ represents the prediction model (which could be pre-trained or learned from the data).

\begin{equation}
	\hat{x}_{T+k} = f(x_1, x_2, ..., x_{T+k-1})
	\label{eq: prediction}
\end{equation}

Assume the actual observed values $X_{actual} = \{x_{T+1}, x_{T+2}, ..., x_{T+\tau}\}$. The predicted values $\hat{X}$ are then compared with $X_{actual}$. An anomaly is flagged if the deviation $\delta$ between $\hat{X}$ and $X_{actual}$ exceeds a predefined threshold $\epsilon$.

Currently, most works under this classification are based on metrics, with only a few focusing on logs. We categorize these works according to the different LLM-based approaches they use into the following groups: \textit{foundational model}, \textit{fine-tuning approach}, \textit{prompt-based approach}, and \textit{embedding-based approach}.

Several works have trained foundation models specifically for time-series forecasting tasks. TimeGPT~\cite{garza2023timegpt} utilizes an encoder-decoder structure with multiple layers to pretrain the first foundation model for time series. Lag-llama~\cite{rasul2023lag} employs a decoder-only transformer architecture that uses lags as covariates to pretrain a general-purpose foundation model for univariate probabilistic time series forecasting. TimesFM~\cite{das2023decoder} leverages a patched-decoder style attention model to pretrain a time-series foundation model for forecasting that performs well across different forecasting history lengths, prediction lengths, and temporal granularities. TTM~\cite{ekambaram2024ttms} utilizes a lightweight TSMixer architecture that incorporates innovations like adaptive patching, diverse resolution sampling, and resolution prefix tuning to handle pre-training on varied dataset resolutions with minimal model capacity.

Although the aforementioned works have developed foundational models for data forecasting tasks, these models often fall short in specific scenarios. Therefore, fine-tuning for particular tasks and contexts is necessary. Zhou et al.~\cite{zhou2023one} utilizes Frozen Pretrained Transformer (FPT) to fine-tune language or computer vision models. LLM4TS~\cite{chang2023llm4ts} adopts a two-stage fine-tuning process: first conducting supervised fine-tuning to orient the LLM towards time-series data, followed by task-specific downstream fine-tuning. Khanal et al.~\cite{khanal2024domain} proposes a one-step fine-tuning method that incorporates a percentage of source domain data into the target domains, providing the model with diverse time series instances and fine-tuning the pre-trained model using a gradual unfreezing technique. LogFiT~\cite{almodovar2024logfit} fine-tunes a BERT-based pre-trained bidirectional encoder representation language model using log data to recognize the language patterns of normal log data. When new log data appears, the model's top-k token prediction accuracy can serve as a threshold for determining whether the new log data deviates from the normal log data.

Training foundational models or fine-tuning requires significant time and hardware resources. Therefore, several works have been proposed to perform predictions without retraining pretrained models. PromptCast~\cite{xue2023promptcast} transforms numerical input and output into prompts, framing the forecasting task in a sentence-to-sentence manner. Time-LLM~\cite{jin2023time} proposes Prompt-as-Prefix (PaP), which enriches the input context and directs the transformation of reprogrammed input patches. The transformed time series patches from the LLM are then projected to obtain the forecasts. TEMPO~\cite{cao2023tempo} decouples complex time series into trend, seasonal, and residual components, mapping them to corresponding latent spaces to create inputs recognizable by GPT. To achieve this, TEMPO constructs a prompt pool and assigns different prompts to different decoupled components, enabling the model to adapt to changes in the time series distribution using historical information. LSTPrompt~\cite{liu2024lstprompt} decomposes the time-series forecasting task into short-term and long-term prediction sub-tasks and uses Chain-of-Thought (CoT) techniques to tailor prompts for each sub-task.

Some approaches rely on the embedding capabilities of large models to perform prediction tasks, particularly in the context of log data. The embedding capabilities of these large models are effective in extracting semantic information from logs. Egil et al.~\cite{karlsen2024large} employs an autoencoding method to compress the embedding representation and conducts anomaly detection through self-supervised learning. Harold et al.~\cite{ott2021robust} utilizes embeddings from models like BERT, GPT-2, and XLNet, chaining the embedding vectors through time in a recurrent neural network (BiLSTM) to learn normal system behavior. LogADSBERT~\cite{hu2023research} leverages the Sentence-BERT model to extract semantic features from log events, followed by the utilization of BiLSTM for log event prediction.

\textbf{Reconstruction-Based Methods.} Reconstruction-based anomaly detection utilize models to reconstruct the input data and detect anomalies by measuring the reconstruction error. If the reconstruction error exceeds a certain threshold, the data point is considered an anomaly. This approach assumes that anomalies cannot be accurately reconstructed by the model.

Formally, assume $X = \{x_1, x_2, ..., x_T\}$ represent the input data, where $x_t$ is the observed value at time $t$. The reconstruction-based method uses a model $g$ to generate a reconstructed sequence $\hat{X} = \{\hat{x}_1, \hat{x}_2, ..., \hat{x}_T\}$, where $\hat{x}_t = g(x_t)$.

Then the reconstruction error $\delta$ is calculated as the difference between the observed value $X$ and the reconstructed value $\hat{X}$, If the reconstruction error $\delta$ exceeds a predefined threshold $\epsilon$, the time window $T$ is flagged as having an anomaly.

Reconstruction-based anomaly detection methods are more commonly used with small-scale models, but they are relatively rare in scenarios involving large language models (LLMs). Currently, existing LLM-related methods are primarily based on masked modeling techniques. LANoBERT~\cite{lee2023lanobert} uses a BERT model to learn through masked language modeling and performs unsupervised anomaly detection during testing by calculating the masked language modeling loss function for each log key. Prog-BERT-LSTM~\cite{shao2022log} proposes a progressive masking strategy to aggregate the text semantic vector and sequence feature vector. SimMTM~\cite{dong2024simmtm} pre-trains deep models based on metrics data by learning to reconstruct the masked content based on the unmasked part, recovering masked time points by the weighted aggregation of multiple neighbors outside the manifold. This approach eases the reconstruction task by assembling ruined but complementary temporal variations from multiple masked series.

\textbf{Classification-Based Methods.} Classification-based anomaly detection classify input data as normal or anomalous based on predefined labels or learned features. Classification-based methods require a labeled dataset for training and typically use supervised learning algorithms to distinguish between normal and anomalous behavior.

Formally, assume that $X = \{x_1, x_2, ..., x_T\}$ represent the input data for a given time window $T$, and $y \in \{0, 1\}$ represents the corresponding label for that time window (0 for normal and 1 for anomalous). The classification process can be represented as Equation~\ref{eq: classification}, where $g$ is the classification model.

\begin{equation}
	\hat{y} = g(X)
	\label{eq: classification}
\end{equation}

During the training phase, the model $g$ is learned using the labeled dataset $(X, y)$. In the detection phase, the model $g$ classifies each time window $X$ into normal or anomalous, and if $g(X) = 1$, the time window $X$ is flagged as an anomaly. Based on the model responsible for the final decision and classification, we categorize classification-based anomaly detection into two types: \textit{LLM-assisted classification} and \textit{LLM-decision classification}.

\textit{LLM-assisted classification} methods utilize large language models to assist smaller-scale models in making final decisions. LLMs provide embeddings or feature extractions, which are then used by smaller, specialized models to classify anomalies.

These works primarily focus on log-based anomaly detection, leveraging LLM embeddings to capture semantic information from log data. RobustLog~\cite{zhang2019robust} was one of the earliest to use a pretrained model (FastText algorithm) to extract semantic information from log events and represent them as semantic vectors, eventually using an attention-based Bi-LSTM model for classification. Building on this, PLELog~\cite{yang2021semi} used a similar semantic information extraction algorithm as RobustLog and transformed the classification-based algorithm into a semi-supervised one. NeuralLog~\cite{le2021log} employed Word2Vec for semantic information transformation and eliminated log parsing to address errors caused by parsing mistakes. Although RobustLog, PLELog, and NeuralLog considered extracting semantic information from logs, the models and algorithms they used for semantic extraction cannot be classified as LLMs.

Building on these, Egil et al.~\cite{karlsen2023exploring} used BERT for semantic information extraction, demonstrating that using LLMs for semantic extraction yields better detection results. MultiLog~\cite{zhang2024multivariate, kang2022separation, zhang2024time, zhang2021two} used BERT to extract semantic information from each node in distributed software and performed semantic compression, ultimately using an AutoEncoder at the coordinator node for fusion and classification. LogST~\cite{zhang2022logst} used SBERT to extract the semantics of log events and finally employed a GRU model for anomaly detection. Ji et al.~\cite{ji2023log} combined SBERT for extracting log embedding information with GPT-2 for semantic transformation, ultimately using an alarm strategy layer for anomaly detection.

\textit{LLM-decision classification} methods utilize large language models directly for decision-making. The LLMs are fine-tuned or prompted to classify input data as normal or anomalous without relying on smaller models for the final classification. These methods can be divided into two categories: training-based methods and prompt-based methods.

Several training-based works utilize BERT and its variants, predominantly focusing on log data. BERT-Log~\cite{chen2022bert} employs a pretrained language model to learn the semantic representation of normal and anomalous logs and fine-tunes the BERT model using a fully connected neural network to detect anomalies. LogPrompt~\cite{zhang2023logprompt} leverages prompts to guide the pretrained language model to better understand the semantic and sequential information of logs, avoiding the need to train a model from scratch. HilBERT~\cite{huang2023improving} uses log template information from 17 log datasets on loghub~\cite{he2023loghub} for full fine-tuning in anomaly detection tasks. To reduce the high cost of full fine-tuning, LogBP-LORA~\cite{he2023parameter} proposes a parameter-efficient log anomaly detection scheme based on BERT and Low-Rank Adaptation (LoRA), which introduces bypass weight matrices and updates only bypass parameters instead of all original parameters.

There are also some works utilize metrics data for fine-tuning. TS-Bert~\cite{dang2021ts} performs full fine-tuning on BERT using metrics data. Zhou et al.~\cite{zhou2023one} fine-tune pre-trained models like GPT-2 and BERT using layer-freezing methods. Additionally, some works use knowledge distillation to learn anomaly detection knowledge from LLMs. AnomalyLLM~\cite{liu2024large} trains a student model for metrics-based anomaly detection using GPT-2 as the teacher model.

However, more novel approaches avoid training models and instead leverage prompt engineering to guide LLMs in performing classification tasks. As large language models grow in scale, it is often believed that they inherently possess anomaly detection capabilities, which can be harnessed through effective prompts. Some methods simply input prompts to the LLM to determine the presence of anomalies.

Many works focus on log data. Priyanka et al.~\cite{mudgal2023assessment} conduct a study on zero-shot log-based anomaly detection using ChatGPT, indicating that the current version of ChatGPT has limited performance in this task. Chris et al.~\cite{egersdoerfer2023early} not only prompt ChatGPT to detect anomalies in log segments but also employ input summary techniques combined with historical data for a more comprehensive assessment, demonstrating better results. RAGLog~\cite{pan2023raglog} enhances LLM-based anomaly detection by integrating retrieval-augmented generation techniques with vector databases, improving detection accuracy. LogGPT~\cite{qi2023loggpt} employs chain-of-thought (CoT) prompting to strengthen the LLM's performance in anomaly detection, enabling a more nuanced and effective analysis.

Unlike log data-based approaches, metrics-based works often require special prompt embedding for metrics data. TEST~\cite{sun2023test} first tokenizes metrics data, builds an encoder to embed them by instance-wise, feature-wise, and text-prototype-aligned contrast, and then creates prompts to make the LLM more receptive to embeddings, finally implementing metrics-based classification. Sigllm~\cite{alnegheimish2024large} employs a time-series-to-text conversion module, as well as end-to-end pipelines that prompt language models to perform time series anomaly detection, directly asking the language model to indicate which elements of the input are anomalies. TabLLM~\cite{hegselmann2023tabllm} prompts large language models with a serialization of the metrics data into a natural-language string for few-shot and zero-shot metrics data classification.

\section{Root Cause Analysis}
\label{sec:root_cause_analysis}

Once an anomaly is detected in the software system, it is necessary to further locate and classify the anomaly. Depending on the specific tasks performed, root cause analysis can be categorized into failure localization (Section~\blueref{sec:failure_localization}), failure category classification (Section~\blueref{sec:failure_category_classification}), and root cause report generation (Section~\blueref{sec:root_cause_report_generation}).

\begin{figure}[h]
	\centering
	\includegraphics[width=\textwidth]{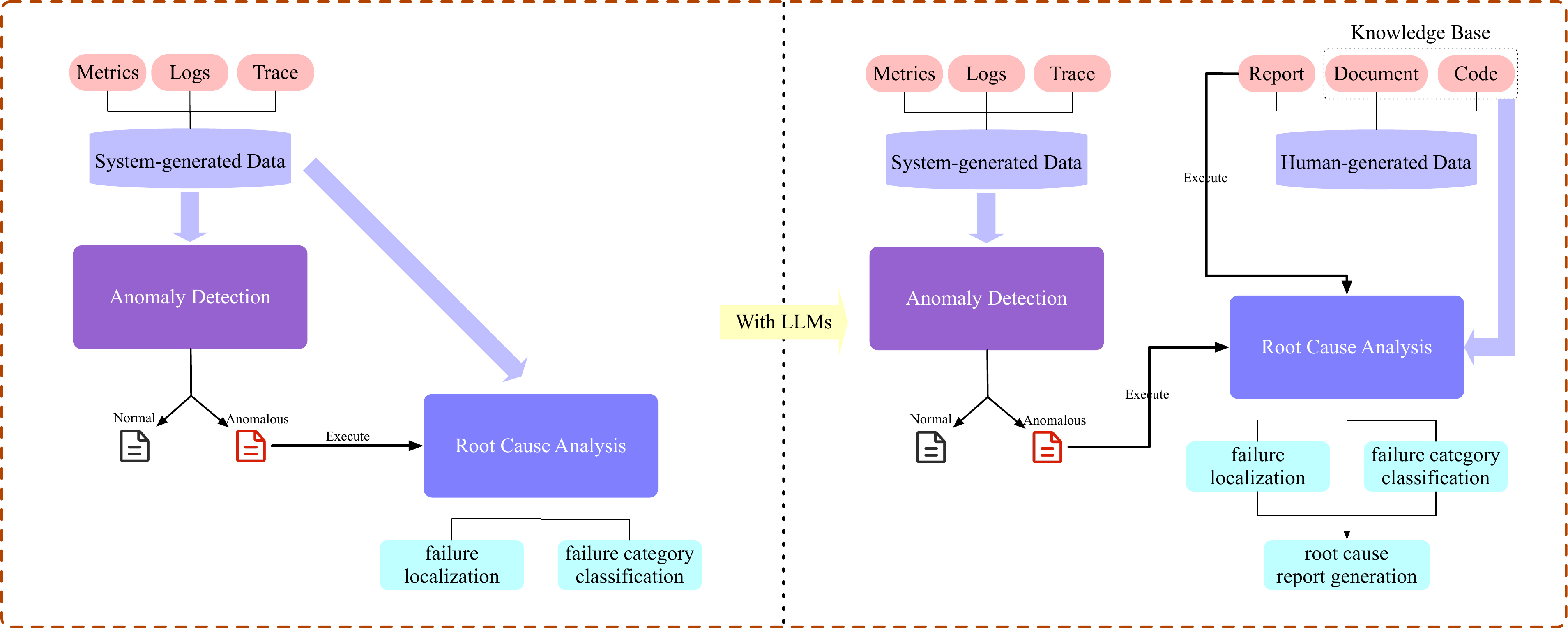}
	\caption{Evolution of Root Cause Analysis with the Rise of LLMs}
	\label{fig: root-cause-analysis}
\end{figure}

As illustrated in Figure~\ref{fig: root-cause-analysis}, before the rise of large language models, these tasks typically relied on system-generated data and utilized automated failure perception methods to identify anomalies, which was then used for failure localization and failure category classification. However, with the advent of large language models, the starting point for root cause analysis has shifted from automated failure perception to user-generated data, particularly incident reports which has introduced the ability to analyze natural language. Additionally, root cause analysis now incorporates other human-generated data, such as documentation and code, as supplementary knowledge sources to enhance the analysis. Furthermore, based on the natural language understanding and generation capabilities of LLMs, it is now possible to bypass the processes of failure localization and failure category classification, and directly generate root cause reports.

\subsection{Failure Localization}
\label{sec:failure_localization}

Failure localization aims to identify the specific component or machine where the anomaly occurred, often using methods such as causal discovery. In microservices environments, it can pinpoint the exact service or machine experiencing the issue. Additionally, it may involve identifying the specific log or metric entry that marked the onset of the anomaly. While widely used in traditional, non-LLM-based works, this type of approach is less common in current LLM-based methods.

Formally, assume $S = \{s_1, s_2, ..., s_n\}$ represent the set of all components or services in the system, and let $L = \{l_1, l_2, ..., l_m\}$ represent the set of log entries or metric data points. The goal of failure localization is to identify the subset $S' \subseteq S$ and $L' \subseteq L$ where the anomalies are most likely to have originated. This process can be represented as Equation~\ref{eq: failure-localization}, where $A$ represents the observed anomalies, and $P(S', L' \mid A)$ is the probability that the components in $S'$ and log entries or metrics in $L'$ are responsible for the observed anomalies. 

\begin{equation}
	(S', L') = \mathop{\arg\max}\limits_{S', L'} P(S', L' \mid A)
	\label{eq: failure-localization}
\end{equation}

In the era of large language models (LLMs), this type of work is relatively rare, and the identified failure types to be located are more diverse. Komal et al.~\cite{sarda2023adarma, sarda2023leveraging} leverage zero-shot prompt engineering approaches based on metrics data to identify problematic host names, and deployment names. Devjeet et al.~\cite{roy2024exploring} utilize incident reports with React~\cite{yao2022react} for root cause analysis in an out-of-domain setting on a static dataset of real-world production incidents, ultimately identifying the problematic task names. RealTCD~\cite{li2024llm} leverages domain knowledge to discover temporal causal relationships for root cause analysis without interventional targets, introducing LLM-guided meta-initialization to extract meta-knowledge from textual information hidden in systems to enhance discovery quality. LogConfigLocalizer~\cite{shan2024face} proposes an LLM-based two-stage strategy for end-users to localize root-cause configuration properties based on logs. mABC~\cite{zhang2024mabc} utilizes alert data (a processed form of metrics data) and employs a multi-agent blockchain-inspired collaboration to identify the specific node where the anomaly occurred.

\subsection{Failure Category Classification}
\label{sec:failure_category_classification}

This approach identifies the type of anomaly the system is experiencing, such as CPU resource shortages, memory shortages, or software configuration errors. The advent of large language models (LLMs) has enhanced the cross-platform generality of these methods and expanded their ability to classify a wider range of anomalies.

Formally, assume $S = \{s_1, s_2, ..., s_n\}$ represents the set of input data items, where each input item $S_i$ describes an observed anomaly. The goal of anomaly classification is to assign each input item $S_i$ a label $l_k$ from a predefined set of anomaly classes $L = \{l_1, l_2, ..., l_m\}$. This process can be represented as Equation~\ref{eq: failure-localization}, where $f$ is a classification function that maps each input item $s_i$ to an anomaly class $l_k$.

\begin{equation}
	f: S \rightarrow L \quad \text{such that} \quad f(s_i) = l_k
	\label{eq: failure-category-classification}
\end{equation}

In the era of large language models, failure category classification methods can be divided into two categories: \textbf{Fine-Tuning-Based Methods} and \textbf{Prompt-Based Methods}.

\textbf{Fine-Tuning-Based Methods.} A considerable number of failure category classification methods are based on fine-tuning and rely on system-generated data. BERTOps~\cite{gupta2023learning} applies full fine-tuning on BERT using log data from 12 Loghub datasets~\cite{he2023loghub} and 5 proprietary data sources, achieving state-of-the-art results in failure category classification. Zhou et al.~\cite{zhou2023one} fine-tune pre-trained models like GPT-2 and BERT using layer-freezing methods based on metrics data, effectively performing failure category classification tasks.

\textbf{Prompt-Based Methods.} Due to the enhanced natural language understanding capabilities of larger pre-trained models, prompt-based methods have become increasingly prevalent and have even become the mainstream approach for failure category classification. Some of these methods still primarily rely on system-generated data. Andres~\cite{quan2023heterogeneous} uses large language models as a message classifier to perform failure category classification on log data. TEST~\cite{sun2023test} builds an encoder based on metrics data to embed them by instance-wise, feature-wise, and text-prototype-aligned contrast, and then creates prompts to make the LLM more receptive to embeddings, achieving excellent results in failure category classification.

Other methods incorporate both system-generated data and human-generated data. D-Bot~\cite{zhou2023d} and LLMDB~\cite{zhou2024llm} use metrics data as a foundation and employ knowledge augmentation techniques (including retrieval-augmented generation and tool-augmented generation). They also integrate human-generated data such as technical manuals, maintenance documents, and API descriptions as additional knowledge sources to classify database failures.

Additionally, some methods leverage mainly on human-generated data, especially incident reports, for classification. PACE~\cite{zhang2023pace} utilizes a two-stage approach: initially, it evaluates its confidence based on historical incident data using vector retrieval methods, considering its assessment of the evidence strength. Subsequently, it uses an LLM to review the root cause generated by the predictor. RCACopilot~\cite{chen2024automatic} synthesizes logs, metrics, and trace data to generate corresponding incident reports and employs a combination of retrieval-augmented generation (RAG) and in-context learning to predict the anomaly's root cause category, providing an explanatory narrative.

\subsection{Root Cause Report Generation}
\label{sec:root_cause_report_generation}

With the enhanced natural language generation and reasoning capabilities of large language models, cutting-edge research is moving beyond isolated tasks of failure location and failure category classification. Instead, there is a growing focus on creating comprehensive and more easily understandable root cause reports. These reports not only include information on failure location and failure category classification but also provide detailed reasoning about the cause of the failure, which significantly aids system maintenance personnel in analyzing and resolving issues.

Toufique et al.~\cite{ahmed2023recommending} conduct the first large-scale empirical study to evaluate the effectiveness of large language models (LLMs) in assisting engineers with root cause analysis and mitigation of production incidents. Their study, performed at Microsoft, involves over 40,000 incident reports and compares several LLMs in zero-shot, fine-tuned, and multi-task settings using semantic and lexical metrics. They utilize various natural language generation evaluation metrics (e.g., BLEU-4, NUBIA) to evaluate the generated root cause reports, demonstrating the efficacy and future potential of using LLMs for resolving cloud incidents.

Building on the study results of Toufique et al.~\cite{ahmed2023recommending}, many subsequent works have focused on optimizing LLM-generated root cause reports using various prompt engineering techniques. Zhang et al.~\cite{zhang2024automated} propose an in-context learning approach based on GPT-4, which eliminates the need for fine-tuning. Drishti et al.~\cite{goel2024x} leverage data from different stages of the software development lifecycle (SDLC) and use a method called In-Context Examples Retrieval, a retrieval-augmented generation (RAG) approach, to enhance the LLM's ability to generate root cause reports. RCAgent~\cite{wang2023rcagent} employs tool-augmented techniques to improve LLM-generated root cause reports, incorporating a unique Self-Consistency for action trajectories, as well as a suite of methods for context management, stabilization, and importing domain knowledge.

\section{Auto Remediation}
\label{sec:auto_remediation}

Once the root cause of the software failure has been identified, the next step is remediation. In the process of software maintenance, there are four main entities involved: the software system, failure incident, on-call engineer (OCE), and large language model (LLM). During the operation of the software system, failure incidents may occur. OCEs and LLMs collaborate to analyze these failure incidents and ultimately complete the remediation.

\begin{figure}[h]
	\centering
	\includegraphics[width=\textwidth]{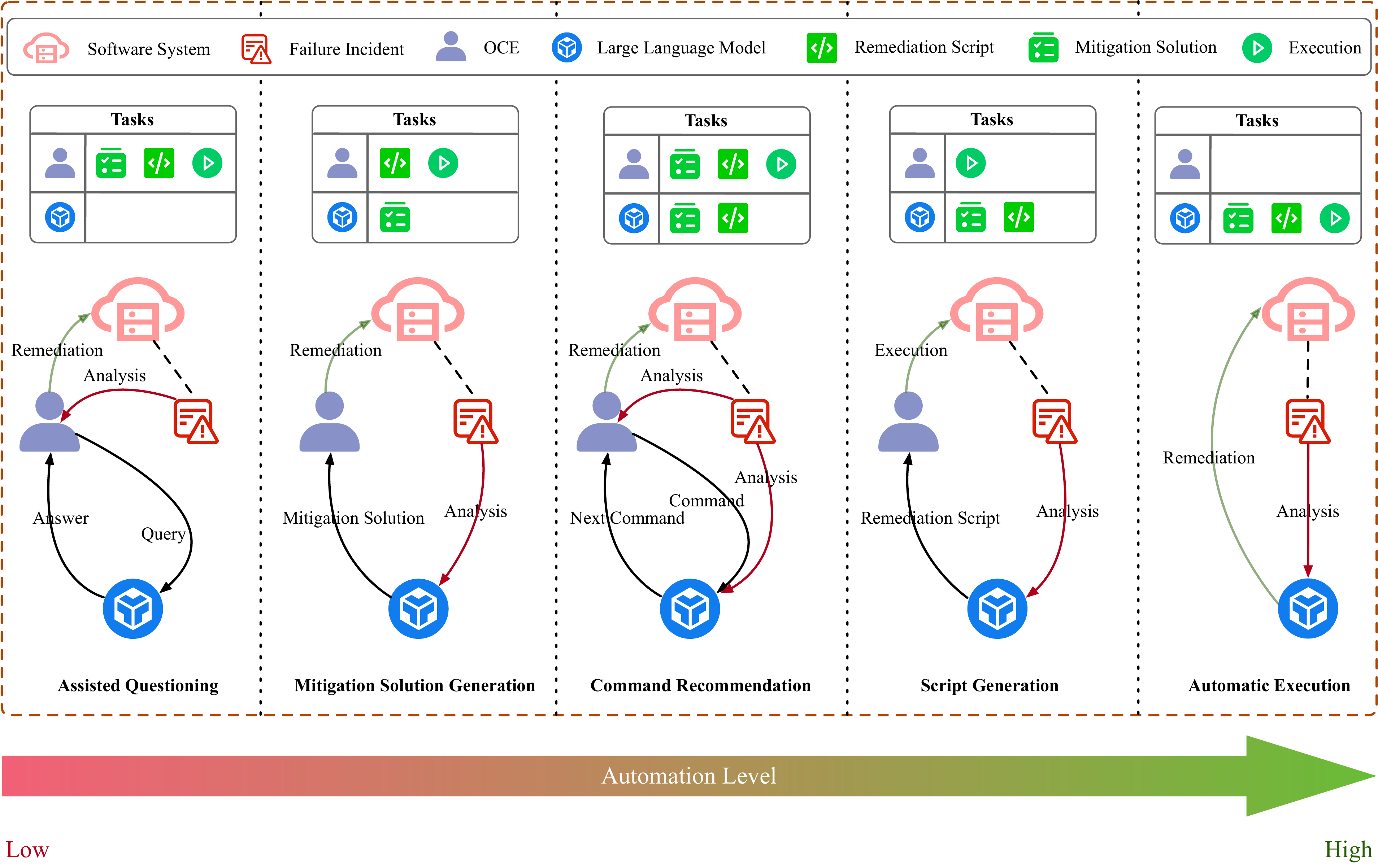}
	\caption{Various Types of Auto Remediation Approaches}
	\label{fig: auto-remediation}
\end{figure}

As shown in Figure~\ref{fig: auto-remediation}, this survey categorizes auto remediation approaches into five types based on the level of automation from low to high: Assisted Questioning(Section~\blueref{sec:assisted_questioning}), Mitigation Solution Generation(Section~\blueref{sec:mitigation_solution_generation}), Command Recommendation(Section~\blueref{sec:script_recommendation}), Script Generation(Section~\blueref{sec:script_generation}), and Automatic Execution(Section~\blueref{sec:automatic_execution}).

\subsection{Assisted Questioning}
\label{sec:assisted_questioning}

Assisted questioning methods for auto remediation have the lowest level of automation but offer the most flexibility. These methods function as specialized question-answering models in the AIOps domain, allowing OCEs to pose any questions to the model, which then provides specialized responses. The model can even utilizes tools to query the software system for necessary metrics. This approach naturally aligns with the natural language processing and reasoning capabilities of large language models. Currently, these methods can be divided into two categories: \textbf{Fine-Tuning-Based Methods} and \textbf{Prompt-Based Methods}.

\textbf{Fine-Tuning-Based Methods.} OWL~\cite{guo2023owl} utilizes supervised fine-tuning and mixture adapter tuning methods to train a set of large language models for knowledge querying based on the LLaMA model and their OWL-Instruct dataset. Gijun et al.~\cite{park2023formulating} propose an interactive AI assistant to aid IT operators in managing deployments. This assistant offers insights into microservice behavior, performance bottlenecks, and preemptive issue resolution, ensuring smooth operations based on a fine-tuned kullm-polyglot-12.8b-v2 model.

\textbf{Prompt-Based Methods.} DB-GPT~\cite{xue2023db} integrates large language models with traditional database systems to enhance user experience and accessibility. It is designed to understand natural language queries and provide context-aware responses. DB-GPT includes a novel retrieval-augmented generation (RAG) knowledge system, an adaptive learning mechanism that continuously improves performance based on user feedback, and a service-oriented multi-model framework (SMMF) with powerful data-driven agents. OpsEval~\cite{liu2023opseval} presents a comprehensive task-oriented AIOps benchmark designed for LLMs, demonstrating how various LLM techniques—such as zero-shot, chain-of-thought, and few-shot in-context learning—can affect the performance of AIOps.

\subsection{Mitigation Solution Generation}
\label{sec:mitigation_solution_generation}

Mitigation solution generation has a higher level of automation compared to assisted questioning. These methods use large language models to analyze failure incidents and ultimately produce mitigation solutions for OCEs to reference in remediating the software system. This step is often linked with root cause report generation; that is, while generating the root cause report, the LLM can also directly generate the mitigation solution.

As mentioned in Section~\ref{sec:root_cause_report_generation}, Toufique et al.~\cite{ahmed2023recommending} conduct the first large-scale empirical study to evaluate the effectiveness of large language models (LLMs) in assisting engineers with root cause analysis and mitigation of production incidents. Drishti et al.~\cite{goel2024x} leverage data from different stages of the software development lifecycle and utilize an in-context examples retrieval and retrieval-augmented generation approach to enhance the LLM's ability to generate mitigation solutions. Pouya et al.~\cite{hamadanian2023holistic} propose a framework analogous to an OCE’s natural thought process, which includes three modules as LLM agents: hypothesis former, hypothesis tester, and mitigation planner. Wang et al.~\cite{wang2023network} propose a control policy generation framework based on a language model for network traffic (NetLM) to refine intent incrementally through different abstraction levels.

\subsection{Command Recommendation}
\label{sec:script_recommendation}

In mitigation solution generation, the next step is to create corresponding remediation scripts based on the root cause. The first step is to assist OCEs by recommending commands. This commonly involves two approaches: OCEs input the previous command, and the system recommends the next command; or OCEs input the task to be completed, and the system recommends a command that can accomplish that task.

ShellGPT~\cite{shi2023shellgpt} belongs to the first approach. It is based on the GPT series models, trained on a corpus that aligns shell language with natural language. It is fine-tuned for command recommendation tasks related to shell language understanding, involving the recommendation of the most appropriate shell command given a sequence of commands. DB-GPT~\cite{xue2023db} falls under the second approach. It employs a retrieval-augmented generation method to understand natural language queries, provide context-aware responses, and generate precise and complex SQL queries.

\subsection{Script Generation}
\label{sec:script_generation}

The aforementioned command recommendation methods still require OCEs to write some commands or input natural language instructions, which involving a certain level of human effort. Therefore, many works have begun to directly research the generation of complete remediation scripts for OCEs to execute.

Xpert~\cite{jiang2024xpert} leverages historical incident data and large language models to generate customized Kusto Query Language (KQL) queries for incident management at Microsoft. The framework efficiently extracts common patterns, such as tables and templates, from historical incidents, facilitating effective automation. To address the limitations of traditional natural language processing (NLP) metrics in evaluating domain-specific queries, Xpert incorporates a novel performance metric called Xcore. This tailored metric allows for more comprehensive evaluation from three different perspectives, enhancing the overall quality assessment of the generated KQL queries. Komal et al.~\cite{sarda2023adarma, sarda2023leveraging} is another representative example. They claim to leverage zero-shot prompt engineering approaches based on metrics data to identify root causes. Based on these root causes, they emphasize the value and significance of using code-generation models (including LLaMA, Codex, and GPT-series models) for auto-remediation in self-adaptive microservice architectures (SMA).

A significant body of research has also concentrated on script generation for failure remediation from the perspective of service composition. Pioneering efforts by Marco et al.~\cite{aiello2023service} are among the first to investigate the potential of LLMs in program generation and its implications for service-oriented computing. Complementary to this, Pesl et al.~\cite{pesl2023uncovering} conduct an empirical study identifying six salient scenarios of service compositions from extant literature, experimenting with ChatGPT and GPT-4 as notable implementations of LLM technology. ChatOps4Msa~\cite{wang2023low}, another innovative contribution, harnesses GPT-3.5 to devise a ChatOps mechanism customized for microservices architectures. This mechanism employs ChatOps Query Language (CQL), leveraging a low-code syntax proposed in their research, to compose configuration files. Consequently, it generates complete script code that facilitates service restoration, thereby empowering OCEs in their maintenance tasks.

\subsection{Automatic Execution}
\label{sec:automatic_execution}

Whether it involves command recommendation or script generation, these methods typically require OCEs to manually execute the generated scripts. The ideal AIOps solution would encompass the entire process: automatic analysis, script generation, and automatic execution, thereby eliminating the need for human intervention in software system maintenance. There have been some pioneering efforts in this direction leveraging large language models to achieve a fully automated maintenance workflow.

Charles et al.~\cite{cao4741492managing} delve into the application of LLM-driven AI agents for automating server administration chores in Linux ecosystems. Their empirical investigation showcases a GPT-based AI agent competently performing 150 distinct tasks spanning nine categories, from file manipulation to programming compilations, illustrating the agent’s aptitude for autonomous task execution and adaptability to feedback. This study underscores the potential of LLMs in democratizing intricate server management tasks for users with diverse technical proficiencies. Pitikorn et al.~\cite{khlaisamniang2023generative}, meanwhile, put forward solutions integrating generative AI for failure perception, code generation, debugging, and the automatic generation of reports within self-healing systems. Notably, their work also capitalizes on GPT-4 to devise a comprehensive and efficacious Python code completion mechanism. This enhancement not only bolsters the functionality of backend systems but also facilitates the repair of malfunctioning components, further demonstrating the versatility of advanced LLMs in IT infrastructure maintenance and optimization.

\section{Challenges and Future Directions}
\label{sec:challenges}

Although detailed introductions to the various LLM-based methods in AIOps for failure management have been provided in previous sections, numerous challenges remain in this field. These include computational efficiency and cost in the utilization of LLMs, in-depth usage of diverse data sources, generalizability and model adaptability during software evolution, and hallucinations in LLMs. These challenges will be discussed in detail below.

\subsection{Computational Efficiency and Cost}

Large language models require substantial computational resources and energy, leading to high operational costs. The training and deployment of LLMs involve extensive use of GPUs or TPUs, making them less accessible for many organizations. Additionally, the inference costs associated with LLMs can be prohibitive, especially when real-time or near-real-time responses are required. This poses significant hurdles, particularly for small to medium-sized enterprises or in scenarios where computational resources are constrained.

Among the tasks of AIOps for failure management, the computational overhead of LLMs is most critical for data preprocessing and failure perception tasks. These steps theoretically need to run continuously during the operation of a software system and require high real-time performance. For example, if the failure perception time window is set to 10 seconds, meaning an failure perception process is triggered every 10 seconds, the failure perception model must infer results within 1 second to promptly notify OCEs in case of anomalies. Some non-LLM-based approaches have focused on addressing this issue~\cite{jia2021logflash, lin2024fastlogad}, but currently, no LLM-based work has adequately tackled this problem. However, this issue is becoming increasingly significant in the LLM era.

On the other hand, while the higher computational efficiency and cost associated with LLMs may be acceptable for root cause analysis and auto remediation (since these steps do not need to run continuously within the software system), this issue still requires careful consideration. Excessive computational costs could potentially make LLM-based approaches less effective than using small-scale models in combination with OCEs. For example, a well-designed smaller model augmented by human expertise might achieve comparable results at a fraction of the cost and resource consumption, making it a more practical and scalable solution in certain contexts.

In summary, balancing computational cost and performance is crucial for the application of LLMs in AIOps. If the cost of running LLMs outweighs their benefits, organizations may find it more advantageous to leverage traditional models or hybrid approaches that combine the strengths of smaller-scale models and human oversight. Therefore, future research should focus on optimizing the computational efficiency of LLMs for AIOps, possibly by exploring methods that organically combine small-scale models, large language models, and OCEs. This integrated approach could provide a cost-effective and scalable solution while maximizing the strengths of each component.

\subsection{In-depth Usage of More Diverse Data Sources}

Several important data sources have yet to be utilized in LLM-based approaches for AIOps. As illustrated in Section~\ref{sec:data-source}, there are three crucial types of system-generated data in software system runtime: metrics, logs, and traces. While there has been significant progress in utilizing metrics and logs, no current work has effectively incorporated trace data.

Trace data is particularly valuable as it provides insights into the calls between nodes in a software system cluster and the interactions within internal components. The lack of LLM-based approaches using trace data can be attributed to its high complexity and volume, making it challenging to integrate trace data with LLMs. Traces often involve detailed and nested sequences of events that need to be represented in a manner that LLMs can effectively understand and utilize. However, trace data remains a crucial resource for comprehensive system monitoring and failure management. Future research should focus on developing methods to effectively integrate trace data into LLM-based AIOps frameworks.

On the other hand, while many failure perception works have been based on logs, a significant portion of these works utilize pre-trained models like BERT, which are not particularly large in scale~\cite{chen2022bert, zhang2023logprompt, huang2023improving, he2023loghub, he2023parameter}. These approach result in relatively high training costs. Only a small fraction of works have utilized larger models such as GPT-3.5, but these studies are primarily based on simple datasets~\cite{egersdoerfer2023early, pan2023raglog, qi2023loggpt}. For these datasets, failure perception can often be achieved with traditional machine learning models~\cite{landauer2023critical}, which diminishes the perceived superiority of LLM-based methods. Therefore, future research should explore more effective ways to apply logs to LLM-based approaches. One feasible solution could be prompt embedding, which has been widely used in metrics-based work.

Lastly, in LLM-based root cause analysis, many works have used incident reports as the primary data source and achieved excellent results. However, this approach disrupts the automation flow of AIOps for failure management, as root cause analysis should be triggered by failure perception. Therefore, future work should focus on using system-generated data to trigger failure perception, generate corresponding incident reports, and then apply root cause analysis methods to analyze these generated incident reports.

\subsection{Generalizability and Model Adaptability in Software Evolution}

A prominent challenge confronted in conventional ML-based and DL-based AIOps approaches for failure management revolves around the severe limitation of model performance degradation when applied to different software systems or following modifications to the original system. Numerous efforts have attempted to mitigate this issue by leveraging techniques such as meta-learning and online learning, which aim to retrain models effectively with minimal new data, thereby rejuvenating their performance.

In the context of LLM-based AIOps methodologies, which primarily harness the inferential capabilities of large language models, there is an inherent expectation that these approaches should naturally exhibit a high degree of generalizability across platforms and model adaptability amidst software changes. The reasoning behind this expectation lies in the extensive pre-training of LLMs on diverse textual data, theoretically enabling them to understand and adapt to a wide range of software contexts without requiring substantial retraining.

However, despite this theoretical promise, current research in the field reveals a gap in empirical validation. While some studies that have established foundational models have indeed demonstrated cross-platform effectiveness, the majority of works, especially prompt-based work, have not systematically assessed the generalizability and adaptability of their LLM-based AIOps approaches in real-world scenarios of software evolution. This oversight points to a crucial area for future research.

Therefore, future endeavors in LLM-based AIOps must prioritize rigorous testing and validation of these models' ability to generalize across different software systems and their resilience to changes within a system. This could involve designing experiments that systematically introduce variations in software environments and measuring how well the models maintain or regain their efficacy post-change, possibly through incremental fine-tuning strategies or continuous learning paradigms. Moreover, exploring the synergy between advancements in LLM architectures, transfer learning methodologies, and adaptive learning algorithms will be pivotal in realizing the full potential of LLMs in ensuring robust and adaptable AIOps solutions.

\subsection{Hallucinations in LLMs}

A prevalent issue encountered in LLM methodologies, despite their potent inferential prowess, is their propensity to generate inaccurate or deceptive content. This phenomenon, known as hallucination, undermines the reliability of model outputs, since LLMs can generate seemingly plausible yet entirely fabricated data or trends. Such inaccuracies can lead to misinformed decisions and flawed analyses, posing significant risks in operational and strategic planning.

Currently, there are no dedicated efforts within LLM-based AIOps research addressing this critical problem. The implications of hallucination in AIOps are significant, as erroneous insights can exacerbate system issues rather than resolve them. For instance, in failure perception or root cause analysis, hallucinated outputs could mislead OCEs, resulting in incorrect diagnostics and remediation strategies that could potentially disrupt system operations further.

Addressing hallucination requires rigorous measures to identify, quantify, and mitigate these inaccuracies. This may entail developing specialized evaluation metrics tailored to detect inconsistencies and contradictions in LLM-generated outputs. Advanced techniques such as fact-checking integration, adversarial training to challenge model outputs, and reinforcement learning strategies that incentivize factual correctness are potential solutions. Additionally, incorporating human oversight and validation within LLM workflows can further enhance the reliability of outputs and mitigate the risks associated with hallucinations.

\section{Conclusion}

This survey thoroughly delves into the landscape of AIOps for failure management in the era of large language models (LLMs), encompassing a detailed definition of AIOps tasks for failure management, the data sources for AIOps, and the LLM-based approaches adopted for AIOps. Additionally, this survey explores the AIOps subtasks, the specific LLM-based approaches suitable for different AIOps subtasks, and the challenges and future directions of the domain.

Despite the progress made, several areas require further exploration. These include computational efficiency and cost in the utilization of LLMs, in-depth usage of diverse data sources, generalizability and model adaptability during software evolution, and hallucinations in LLMs.

Future research must concentrate on achieving a optimal balance between computational cost and performance in applying LLMs to AIOps tasks. Expanding the exploration of unique data source attributes and devising sophisticated methods for their integration into LLMs is central to enhancing model performance and adaptability. Additionally, prioritizing rigorous testing and validation to assess models' cross-platform generalizability and their resilience against software changes is of utmost importance. Lastly, the development of specialized metrics to identify and rectify hallucinations within LLM-generated outputs for AIOps applications is imperative.

In conclusion, the realm of AIOps for failure management in the era of LLMs is a dynamically advancing field teeming with both promise and obstacles. Its progression is vital to underpinning the stability and reliability of software systems. By tackling the challenges outlined in the survey, LLM-powered AIOps methodologies for failure management are poised to see broader, more impactful real-world implementation, thereby reinforcing the resilience and efficiency of modern software ecosystems.

\begin{acks}
 This work was supported by the National Key R\&D Research Fund of
 China (2021YFF0704202).
\end{acks}

\bibliographystyle{ACM-Reference-Format}
\bibliography{sample-base}

\end{document}